\documentclass[10pt,preprint]{aastex}

\newcommand{\myemail}{keivan.stassun@vanderbilt.edu}
\newcommand{\bbmax}{BB$_{\rm max}$}
\newcommand{\bbmin}{BB$_{\rm min}$}
\newcommand{\ewca}{EW[\ion{Ca}{2}]}

\shorttitle{PMS X-ray and Optical Variability}
\shortauthors{Stassun et al.}

\begin{document}

\slugcomment{To appear in the Astrophysical Journal}

\title{A Simultaneous Optical and X-ray Variability Study of
the Orion Nebula Cluster.\ II.\ A Common Origin in Magnetic
Activity}

\author{Keivan G.\ Stassun\altaffilmark{1},
M.\ van den Berg\altaffilmark{2},
Eric Feigelson\altaffilmark{3}}

\altaffiltext{1}{Department of Physics \& Astronomy,
Vanderbilt University, Nashville, TN 37235; \myemail}
\altaffiltext{2}{Harvard-Smithsonian Center for Astrophysics, 
60 Garden St., Cambridge, MA 02138}
\altaffiltext{3}{Department of Astronomy \& Astrophysics, 
Penn State University, University Park, PA 16802}

\begin{abstract}
We present a statistical analysis of simultaneous optical and X-ray
light curves, spanning 600~ks, for 814 pre--main-sequence (PMS) stars
in the Orion Nebula Cluster. The aim of this study is to establish
the relationship, if any, between the sites of optical and X-ray
variability, and thereby to elucidate the origins of X-ray production
in PMS stars. In a previous paper we showed that optical and X-ray
variability in PMS stars are very rarely time-correlated. Here, using
time-averaged variability indicators to examine the joint occurrences
of optical and X-ray variability, we confirm that the two forms of
variability are not directly causally related. However, a strong and
highly statistically significant correlation is found between optical
variability and X-ray luminosity. As this correlation is found to be
independent of accretion activity, we argue that X-ray production in
PMS stars must instead be intimately connected with the presence and
strength of optically variable, magnetically active surface regions
(i.e.\ spots) on these stars. Moreover, because X-ray variability and
optical variability are rarely time-correlated, we conclude that the
sites of X-ray production are not exclusively co-spatial with these
regions. We argue that solar-analog coronae, heated by topologically
complex fields, can explain these findings.

\end{abstract}


\section{Introduction\label{intro}}
While it is now well established that low-mass pre--main-sequence
(PMS) stars typically have X-ray luminosities 2--4 orders of magnitude
higher than that of the present-day Sun \citep[e.g.][]{preibisch05},
fundamental questions remain as to how these X-rays are
actually produced. Observations of X-ray flares in these stars
indicate solar-analog magnetic reconnection events, but the
absence of an obvious main--sequence-like rotation-activity
relation in PMS stars has so far confounded attempts to
ascribe PMS X-ray production directly to dynamo-generated
fields \citep{flac-basic,feig03,stass04,rebull06}. Accretion
has also been implicated as a possible driver of PMS X-ray
production, but no consensus has emerged on this point as yet
\citep[e.g.][]{kastner02,stelzer04,stass04}. Indeed, there is evidence
that accretion may act to inhibit X-ray production instead of enhance
it \citep[e.g.][]{flac-basic,preibisch05}. These issues are reviewed
more fully in \citet{feig06}.

X-rays from PMS stars are interesting not only because of the mystery
surrounding their origins. If X-rays trace magnetic activity in young
stars as they do in main sequence stars, then X-rays can provide
important clues to a variety of star-formation questions involving
stellar magnetic fields---the evolution of stellar angular momentum
via magnetized winds and magnetic coupling to circumstellar disks,
for example. X-rays may also be central players in the physics of
molecular clounds and protoplanetary disks, driving the photoionization
of circumstellar material \citep[e.g.][]{glassgold00}. Thus,
understanding the physics of X-ray production in PMS stars
may ultimately prove central to our understanding of star- and
planet-formation \citep{feig06}.

To make progress in understanding {\it how} PMS X-rays are made,
it would be valuable to have constraints on {\it where} the X-rays
are coming from. A key limitation in this regard has been the lack of
simultaneous, independent measures of X-ray production and of other
forms of activity, such as optical variability. So motivated, we have
undertaken an extensive variability study of the Orion Nebula Cluster
(ONC) combining a nearly continuous, 13-d {\it Chandra} observation
with simultaneous, multi-wavelength, time-series photometry in
the visible. The resulting database of simultaneous X-ray and
optical light curves for some 800 members of the ONC represents,
by a factor of hundreds, the largest attempt to date to study the
relationship betweeen X-ray and optical variability in PMS stars
\citep[][hereinafter Paper~I]{stass06}.

In Paper~I, we used this database to search for time-correlated
variability in our simultaneous optical and X-ray light curves with
the goal of establishing the frequency with which X-ray and optical
variations are co-temporal, and to thereby constrain the extent to
which the sources of X-ray and optical variability may be co-spatial
or otherwise causally linked. For example, if X-rays are produced near
accretion shocks on classical T Tauri stars (CTTSs; actively accreting
PMS stars), then one might expect that changes in the strength of these
shocks will induce changes in the strength of the X-ray emission, and
that X-ray and optical variability might therefore be correlated in
time. Similarly, if X-rays originate in coronal structures associated
with magnetically active surface regions on these stars (i.e.\ dark
spots), we might then expect that optical and X-ray variability would
be anti-correlated in time. However, we found that such behavior
almost never occurs. Indeed, except for a single possible case of
``white light" flare emission, simultaneous X-ray/optical variations
were not clearly seen in any of the $800+$ stars observed. The upper
limit on the incidence of time-correlated optical/X-ray variability
in our sample is at most $\sim 5\%$; that is, at least $\sim 95\%$
of the stars in our sample exhibit no evidence for time-correlated
optical and X-ray variability (Paper~I). The implication is that the
sites of optical and X-ray variability in PMS stars are not---in the
vast majority of cases---instantaneously one and the same. Similar
findings have now also been reported from a simultaneous optical/X-ray
monitoring survey of Taurus by \citet{audard07}.

Whereas these findings rule out the simple interpretation
of the X-ray and optical variability both dominated by one site of
emission, in this companion paper we examine the possibility of
a more general relationship between optical and X-ray variability
by investigating {\it time-averaged} optical and X-ray variability
indicators (e.g.\ variability amplitudes) and possible correlations
therein. After reviewing the salient contents of our simultaneous
optical/X-ray variability database, and the time-averaged
variability indicators that we use in our analysis (\S\ref{data}), in
\S\ref{corr-general} we perform a multivariate correlation analysis of
these indicators with one another and with other stellar properties
related to energy production (accretion signatures, bolometric
luminosity, X-ray luminosity). We find that time-averaged optical
and X-ray variability are not well correlated with one another,
implying that they are not causally related. Importantly, however, we
do find optical variability to be strongly correlated with X-ray {\it
luminosity}, and that this correlation exists regardless of whether
the stars in question are actively accreting (CTTSs) or non-accreting
(weak-lined T Tauri stars; WTTSs).

We discuss in \S \ref{discussion} the implications of these results.
We argue that accretion is not in general a significant contributor
to the production of X-rays in CTTSs, and that PMS X-ray production
overall is more intimately related to the presence of magnetically
active surface regions (i.e.\ spots), as traced by optical variability,
in both CTTSs and WTTSs. At the same time, the lack of time-correlated
optical and X-ray variability in the overwhelming majority of
our sample (Paper~I) indicates that the X-rays are not produced in
magnetically active surface regions alone. We summarize our conclusions
in \S \ref{conclusions}.

\section{The COUP synoptic X-ray and optical database\label{data}}
A complete description of our database of simultaneous X-ray and
optical light curves is provided in Paper~I. Briefly, the database
comprises 814 PMS members of the ONC within a 17'$\times$17' region
centered on the Trapezium observed by ACIS onboard {\it Chandra} for
13.2 d. These stars were simultaneously observed at $BVRI$ wavelengths
by small telescopes on the ground during the last 7 d of the {\it
Chandra} observation, with a cadence of $\sim 1\; {\rm hr}^{-1}$.

For each star, the database also includes the star's X-ray luminosity,
$L_X$ \citep{getman05}, and a compilation of other physical parameters
from photometric and spectroscopic catalogs already in the literature,
including bolometric luminosity, $L_{\rm bol}$ \citep{hill97}, and
the equivalent width (EW) of the \ion{Ca}{2} infrared triplet from
\citet{hill98}, a measure of mass accretion rate. As in previous
studies \citep[e.g.][Paper I]{flac-basic,stass04,preibisch05}, we
take the ``accretors" to be those stars with \ion{Ca}{2} strongly in
emission, i.e., EW $\le -1$ \AA, and ``non-accretors" to be those
with \ion{Ca}{2} clearly in absorption, i.e., EW $\ge 1$ \AA. Of
the 493 stars in our study sample with \ion{Ca}{2} measurements, 151
stars are accretors and 145 stars are non-accretors, as defined here
(the remainder show indeterminate values close to 0 \AA)\footnote{As a
caveat, we note that \ewca\ is an imperfect indicator of accretion. As
shown by \citet{sicilia05}, some PMS stars with clear accretion
signatures in H$\alpha$ do not manifest this accretion activity
in \ion{Ca}{2}. It is therefore possible or even likely that some
low-level accretors will go undetected in \ion{Ca}{2}. The criterion
that we adopt for identifying the presence of accretion is thus in
practice a conservative one; stars with \ewca\ $\le -1$ \AA\ and \ewca\
$\ge 1$ \AA\ are securely identified as accretors and non-accretors,
while some stars with indeterminate values close to 0 \AA\ may be
weakly accreting objects that will go unidentified as such.}.

The 814 stars in our database span a very large range of apparent
magnitudes and colors, with $7.3 < I < 18.4$ and $-0.1 < (V-I) < 5.7$,
and a correspondingly large range of spectral types and extinctions,
B3 $<$ SpTy $<$ M7 and $0.0 < A_V < 10.8$ \citep{hill97}. This sample
is thus representative of the underlying ONC population; only the very
massive, heavily obscured, and sub-stellar populations are excluded.

These stars exhibit a large variety of behaviors in their optical
and X-ray variability. Figs.\ 1--6 in Paper~I show the optical and
X-ray light curves for a selection of objects as a representative
visual summary of the range and types of variability observed. Below
we describe the statistical indicators that we use to quantify the
time-averaged X-ray and optical variability.

\subsection{Time-averaged variability indicators\label{methods}}

\subsubsection{X-ray variability\label{xvar}}
The COUP database provides two measures of X-ray variability
in the detected X-ray sources. These are discussed in detail by
\citet{getman05}. 

The first measure is a standard one-sided Kolmogorov-Smirnoff (KS)
test, which tests the null hypothesis that the photon arrival times are
distributed uniformly in time. The KS test thus gives the probability
that the source's X-ray flux is consistent with being non-variable
(i.e.\ constant). 
We take sources with KS test probabilities of less than 0.001 as being
definitively X-ray variable, the expected number of false positives
in our sample then being less than 1. Of the 814 sources in our
study sample, 556 (68\%) are X-ray variable at greater than 99.9\%
confidence by this definition.

The second measure of time-averaged X-ray variability is a Bayesian
Block (BB) analysis \citep{scargle98}, which segments each light curve
into the maximum number of time blocks such that the differences in
the mean flux levels of the blocks are statistically significant. The
BB analysis thus yields a robust measure of the maximum and minimum
flux levels in the source's X-ray light curve (\bbmax\ and \bbmin,
respectively). In our analysis, we use the ratio \bbmax/\bbmin\
to quantify the magnitude of variability in the X-ray light curves
of stars in our sample. The X-ray variables in our database have a
range of amplitudes $0.12 < \log($\bbmax/\bbmin$) < 2.89$ (see
Fig.\ \ref{Jhist}). The KS
and BB measures of time-averaged X-ray variability for the stars in
our study sample are reported in Table \ref{var-table}.

\subsubsection{Optical variability\label{optvar}}
To characterize the time-averaged optical variability of
the stars in our sample we use the $J$ statistic described by
\citet{stetson96}. Similar to the standard $\chi^2$ statistic, the
$J$ statistic measures the residuals of the data about the weighted
mean level of the light curve, and compares those residuals to
the measurement uncertainties. However, unlike a standard $\chi^2$
analysis, the Stetson index simultaneously considers the measurements
from light curves in all available photometric bands of a given
source. This is done by pair-wise matching the data points in time
and comparing the signs of the matched residuals.

$$J \equiv \frac{\Sigma_{i=1}^p w_i \, {\rm sgn}(P_i) \sqrt{|P_i|}}
{\Sigma_{i=1}^p w_i}$$

\noindent
where p is the number of pairs of observations of the star, $P_i =
\delta_{j(i)}\delta_{k(i)}$ is the product of the normalized residuals
of the paired measurements, and $w_i$ is the statistical weight
assigned to each pair \citep[see][for details on weights]{stetson96}.
The normalized residual of measurement $i$ in passband $j$ is given by

$$\delta_{j(i)} = \sqrt{\frac{n}{n-1}} \frac{m_i - \overline{m}}{\sigma_i}$$

\noindent
where $n$ is the number of measurements used to determine the mean
magnitude $\overline{m}$, and $\sigma_i$ is the photometric uncertainty
in measurement $m_i$.

The $J$ statistic is more robust than $\chi^2$ against outliers.
As an example, consider a non-variable star with simultaneous light
curves in three passbands, and consider the case where a cosmic ray
hit appears as a discrepantly high data point in one of these three
light curves. This one outlier will increase the value of the $\chi^2$
statistic for that light curve, making it appear variable in one
passband but not in the others. In contrast, this outlier will tend
to decrease the value of the $J$ statistic. Following the procedure
of \citet{stetson96} to make the $J$ statistic even more robust,
we compute the $J$ statistic iteratively, adjusting the weights of
individual data points and re-computing the weighted mean of each
light curve between successive iterations, thereby minimizing the
influence of strong outliers on the $J$ statistic.

To calculate $J$ in practice, the light curves must be binned in time,
which requires a choice of binning timescale. This binning is not in
the sense of combining adjacent data points; rather, time bins are
defined to determine which data points from the multiple light curves
are associated with one another in the pair-wise matching discussed
above. Given the typical sampling and cadence of the optical light
curves (see Paper~I), we adopt time bins of 1~hr.

\begin{figure}[ht]
\epsscale{0.9}
\plotone{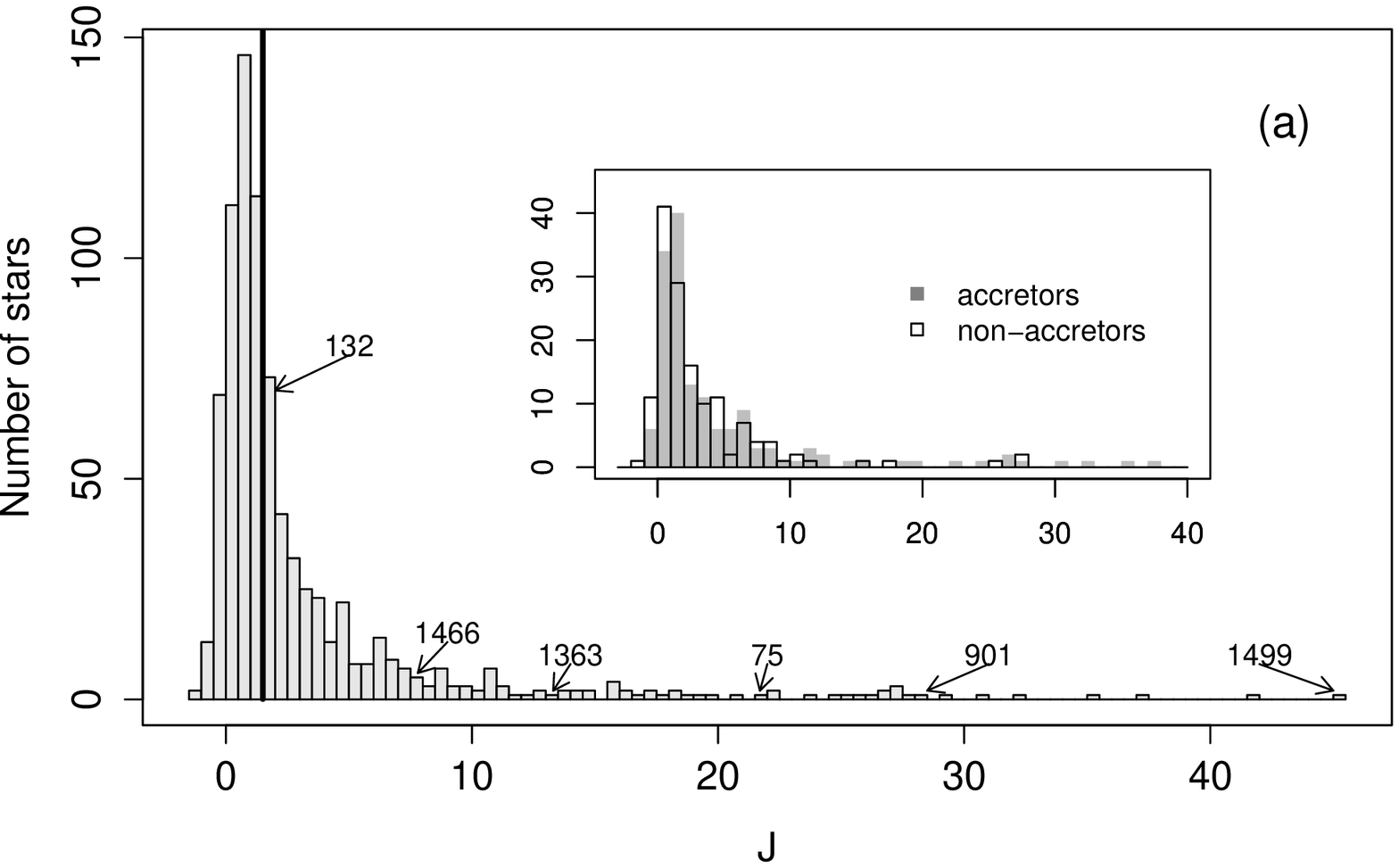}
\plotone{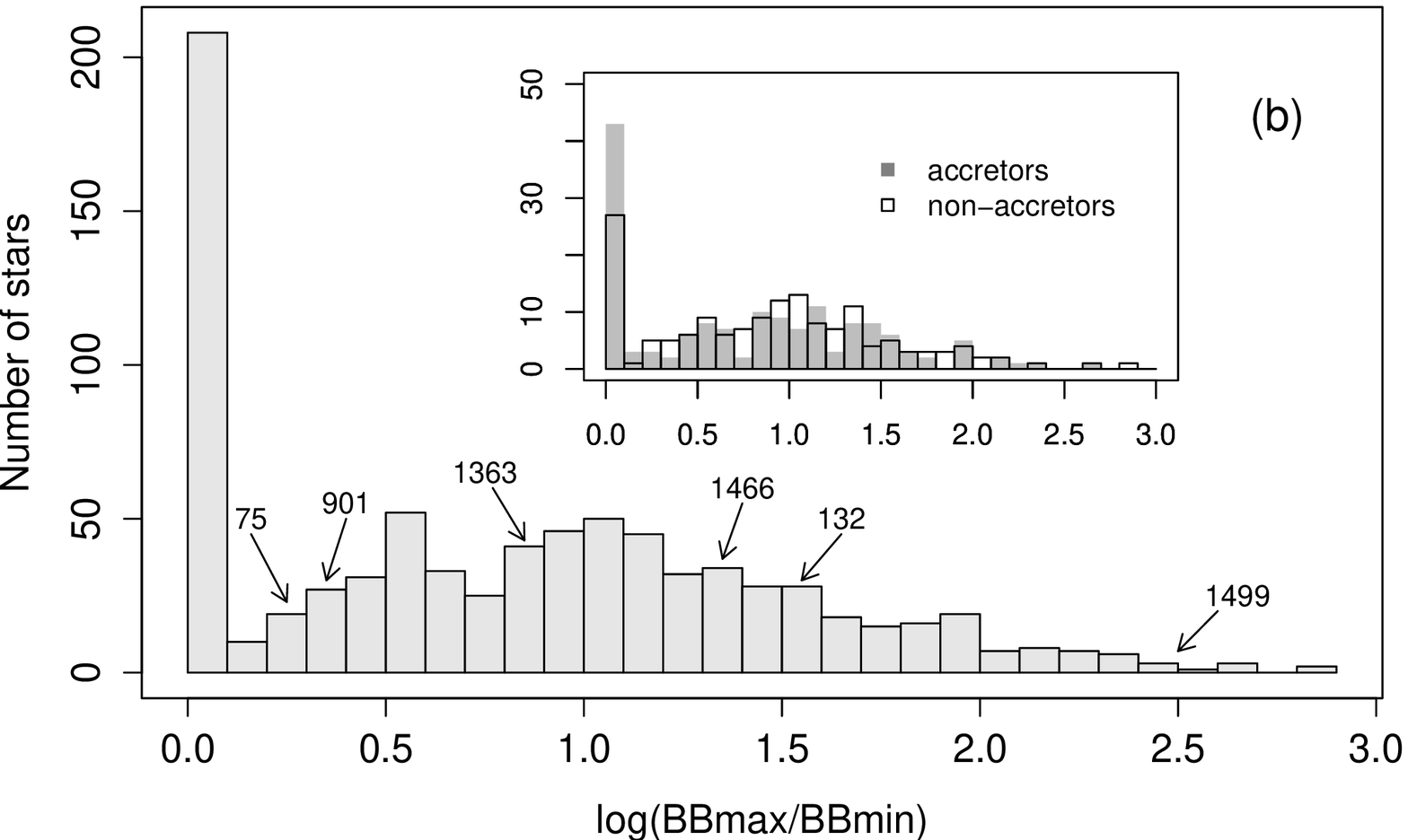}
\caption{\label{Jhist}
(a) Histogram of the $J$ statistic for all stars in the study sample.
Chance correlations in the light curves produce a distribution of positive
and negative $J$ values centered on zero. The most negative value of
$J$ is $-1.50$; thus $J>1.50$ (indicated by the vertical line)
serves as a conservative threshold for identifying stars that are
likely true variables. Stars with various values of $J$ above this
threshold are indicated; their light curves can be found in Paper~I.
Inset panel compares the distribution of $J$ separately for accretors
and non-accretors as defined in \S\ref{data}.
(b) Same as above but for $\log$ \bbmax/\bbmin. The first bin is
dominated by stars with
$\log$ \bbmax/\bbmin $=0$, which are non-variable in X-rays.}
\end{figure}

Values of $J$ for the stars in our study sample are reported in Table
\ref{var-table}. For each star we also report in Table \ref{var-table}
the measures of time-averaged X-ray variability and of accretion
discussed above. The distribution of $J$ for our study sample is shown
in Fig.\ \ref{Jhist}. From this distribution, we can empirically choose
a threshold of $J$, above which we may be confident that the optical
light curves manifest true variability. True variable stars will
show positive values of $J$, the magnitude of which is related to the
amplitude of the variability. In comparison, non-variable stars will
produce a narrow distribution of $J$ centered about zero, with equal
probability for being positive or negative. This narrow distribution
of values about zero represents the probability distribution of $J$
arising from small, chance correlations in the optical light curves,
and we may thus use the most negative value of $J$ as a conservative
estimate of the most positive value of $J$ that will result from such
chance correlations.

The distribution of $J$ in our sample (Fig.\ \ref{Jhist}) shows a
maximal negative value of $J=-1.50$. This value occurs once in our
sample of 814 stars, and we may therefore expect that a value of
$J>1.50$ will occur by chance with a probability of $1/814 \approx
0.001$, which is the same as the conservative criterion we have adopted
for identifying variability in the X-ray light curves. 

With this threshold, 358/814 (44\%) of the stars are optically
variable, with a range $1.5 < J \le 45.5$. The optical and X-ray
light curves for a selection of sources with $J$ spanning this range
are shown in Figs.\ 1--6 in Paper~I. In general, larger values of
$J$ imply greater variability. The threshold of $J>1.50$ that we
adopt for identifying truly variable objects corresponds roughly to
peak-to-peak variability of $\sim 0.05$ mag (typical photometric errors
are 0.01--0.02 mag; Paper~I), though for any particular source the
peak-to-peak variability is different at the different wavelengths
(generally larger at shorter wavelengths due to the increased
temperature contrast). Thus, sources identified as ``non-variable"
in fact represent upper limits of peak-to-peak variability less
than $\sim 5$\%. At the other extreme, objects with the largest $J$
values show peak-to-peak variability in the $V$ band of $\gtrsim 2$
mag. With this range of variability amplitudes, the time-averaged
optical variability of our sample is typical of what has been observed
before in optical variability studies of TTSs \citep[e.g.][]{herbst94}.

\section{Results: Statistical analysis of time-averaged variability 
indicators \label{corr-general}}
Here we report the results of a statistical analysis of these
variability indicators, with the goal of determining their
relationships with one another and with other stellar parameters. We
present three levels of analysis in order of increasing sophistication
(\S\S\ref{var-accretion}--\ref{var-lx}) with a concise summary in
\S\ref{summary}. All statistical analyses reported here were performed
with the {\tt R} software\footnote{{\tt R} is a freely available
language and environment for statistical computing and graphics. See
\url{http://www.r-project.org}.} package.

\subsection{Relation of optical and X-ray variability to accretion
\label{var-accretion}}
One of the aims of our study is to determine whether accretion may be
a significant contributor to the production of X-rays in PMS stars. We
have already seen in Paper~I that there is no evidence in the form of
time-correlated X-ray and optical variability for such a connection,
implying that any causal relationship between accretion and X-ray
production must be indirect. Here we consider whether the presence
of active accretion manifests itself in the time-averaged variability
of our sample.

Applying a Student's $t$ test to the distributions of X-ray variability
(\bbmax/\bbmin) for the accretors and non-accretors, we find a
probability of 38\% that these two groups have similar mean values
of \bbmax/\bbmin. That is, there is not a statistically significant
difference in the level of X-ray variability among the accretors
as compared to the non-accretors. 

In contrast, accretion does manifest itself in the optical
variability of these stars. A $t$ test applied to the same accretor
and non-accretor samples confirms, at the 99\% confidence level, that
accretors exhibit higher average levels of optical variability ($J$)
than the non-accretors. This is as expected, since optical variability
in CTTSs is believed to be driven in part by accretion. This result is
also reassuring, as it demonstrates that the effects of accretion are
discernible in the gross statistical behavior of our study sample, even
though multiple processes may be contributing at once to the observed
variability, and even though the \ewca\ measurements that we use to
identify the accretors were obtained years earlier \citep{hill98}.

\subsection{Joint occurrences of optical and X-ray variability
\label{var-joint}}
Having found that accretion manifests itself in the level of optical
variability of our sample, but not in the level of their X-ray
variability, we next compare the {\it incidences} of optical and
X-ray variability in these stars. We find statistically significant
optical variability among 358/814 (44\%) of the stars (Table
\ref{var-contingency-table}) and statistically significant X-ray
variability among 556/814 (68\%) of the stars (\S\ref{xvar}). A
Pearson $\chi^2$ contingency-table test shows that the probability
that optical and X-ray variability occur with the same frequency
is less than $10^{-6}$. In other words, a statistically significant
fraction of stars that exhibit X-ray variability do not also exhibit
optical variability in our data.

Conversely, we consider whether optically variable stars always
exhibit X-ray variability. Of the 358 stars in our sample that we have
identified as optically variable, 278 (78\%) of them also exhibit
X-ray variability (see Table \ref{var-contingency-table}). The
likelihood that this fraction is consistent with 100\% is again
less than $10^{-6}$. We can therefore state that, to the sensitivity
available, the cause of optical variability cannot {\it at the same
time} be the cause of X-ray variability for a significant fraction
of the 814 observed stars. The two forms of variability simply do
not always occur together during the 7 days of overlap between the
optical and {\it Chandra} observations.

Perhaps the occurrences of optical and X-ray variability are more
strongly related among those stars that are accreting, as one might
expect if active accretion drives both forms of variability. To test
this, we again segregate the stars in our sample into accretors and
non-accretors, as defined earlier. Among the accretors, we find
simultaneous occurrences of optical and X-ray variability for 60
stars (40\%), whereas for the non-accretors we find simultaneous
occurrences of optical and X-ray variability for 64 stars (44\%)
(see Table \ref{var-contingency-table}). This small difference is
not statistically significant (52\% probability that they are not
different) according to a Pearson $\chi^2$ contingency-table test. 

While we find clear evidence that optical and X-ray variability are
not always linked to one another, and that the presence of active
accretion does not increase the likelihood for an association between
the two, this does not necessarily imply that the occurrences of
optical and X-ray variability are completely independent of one
another. In fact, we find that stars showing one form of variability
are more likely to also show variability in the other. For example,
whereas optical variability is found in 44\% (358/814) of the overall
study sample, this fraction increases to 50\% among the stars in our
sample that are X-ray variable (278/556), and 69\% (178/258) of the
X-ray non-variable stars are also optically non-variable (see Table
\ref{var-contingency-table}). A Pearson $\chi^2$ contingency-table test
shows that the likelihood of the two measures of variability being
independent of one another is less than $10^{-6}$. Thus, optical and
X-ray variability are, in fact, related. However, this relationship
is not directly causal, as we now show.

\subsection{Strength of optical and X-ray variability and their
relation to X-ray luminosity\label{var-lx}}
\begin{figure}[ht]
\epsscale{1.0}
\plotone{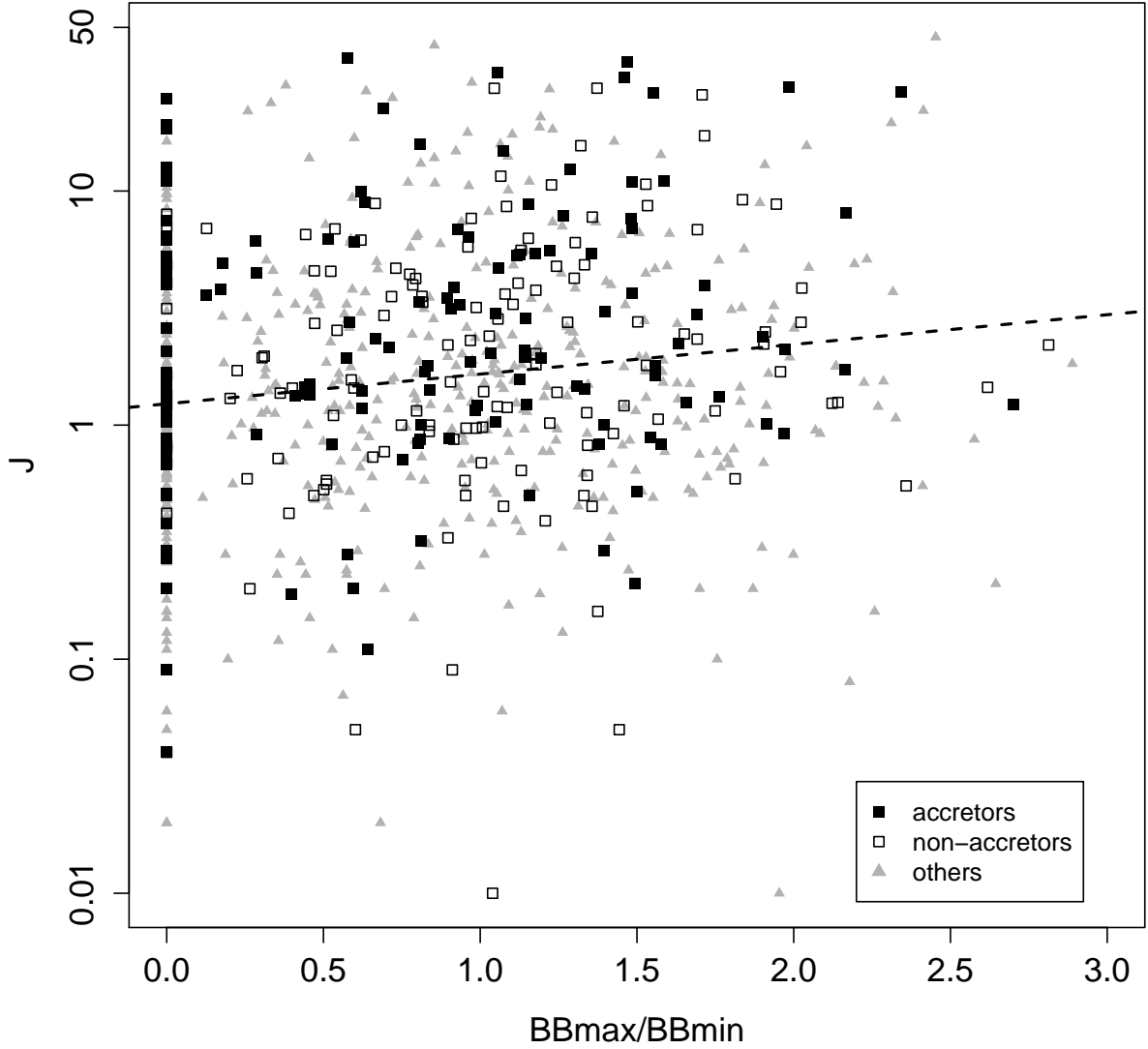}
\caption{\label{Jvar-xvar}
Strength of optical variability as measured by the $J$ statistic is 
plotted vs.\ strength of X-ray variability as measured by the BB
analysis. Stars with negative $J$ have been excluded from the plot
so that it may be displayed on a logarithmic scale (see Fig.\ \ref{Jhist}a
for explanation of negative $J$ values). 
The dashed line shows a linear regression fit to all of
the data. A statistically significant correlation in the sense of the
line is confirmed by a Kendall's $\tau$ test. This correlation is
found to be the result of the mutual correlation of $J$ and \bbmax/\bbmin\
with X-ray luminosity; see text.}
\end{figure}

To explore the possible link between optical and X-ray variability
further, we next examine how the {\it strength} of one may be related
to the other. In Fig.\ \ref{Jvar-xvar}, we plot the strength of optical
variability, as measured by the $J$ statistic (\S \ref{optvar}),
versus the strength of X-ray variability, as measured by the BB
analysis (\S \ref{xvar}). A Kendall's $\tau$ test shows that the two
measures are correlated at a high level of statistical significance,
with the probability that they are not correlated being $3\times
10^{-6}$. The same test conducted separately among the smaller
sub-samples of accretors and non-accretors shows a correlation in the
same sense, albeit at marginal levels of significance (Kendall's $\tau$
probabilities of 0.08 and 0.009 for the accretors and non-accretors,
respectively). Thus, we find that not only does the {\it presence}
of one form of variability correlate with the other (see above),
but that optical and X-ray variability are furthermore correlated in
{\it strength}.

But do these statistical correlations between optical variability
and X-ray variability imply a causal relationship between the two?
The data suggest that this is {\it not} the case, and that the
statistical relationship between optical and X-ray variability is
in fact due to a mutual correlation---the first physical, the second
observational---with X-ray {\it luminosity}.

\begin{figure}[ht]
\epsscale{0.85}
\plotone{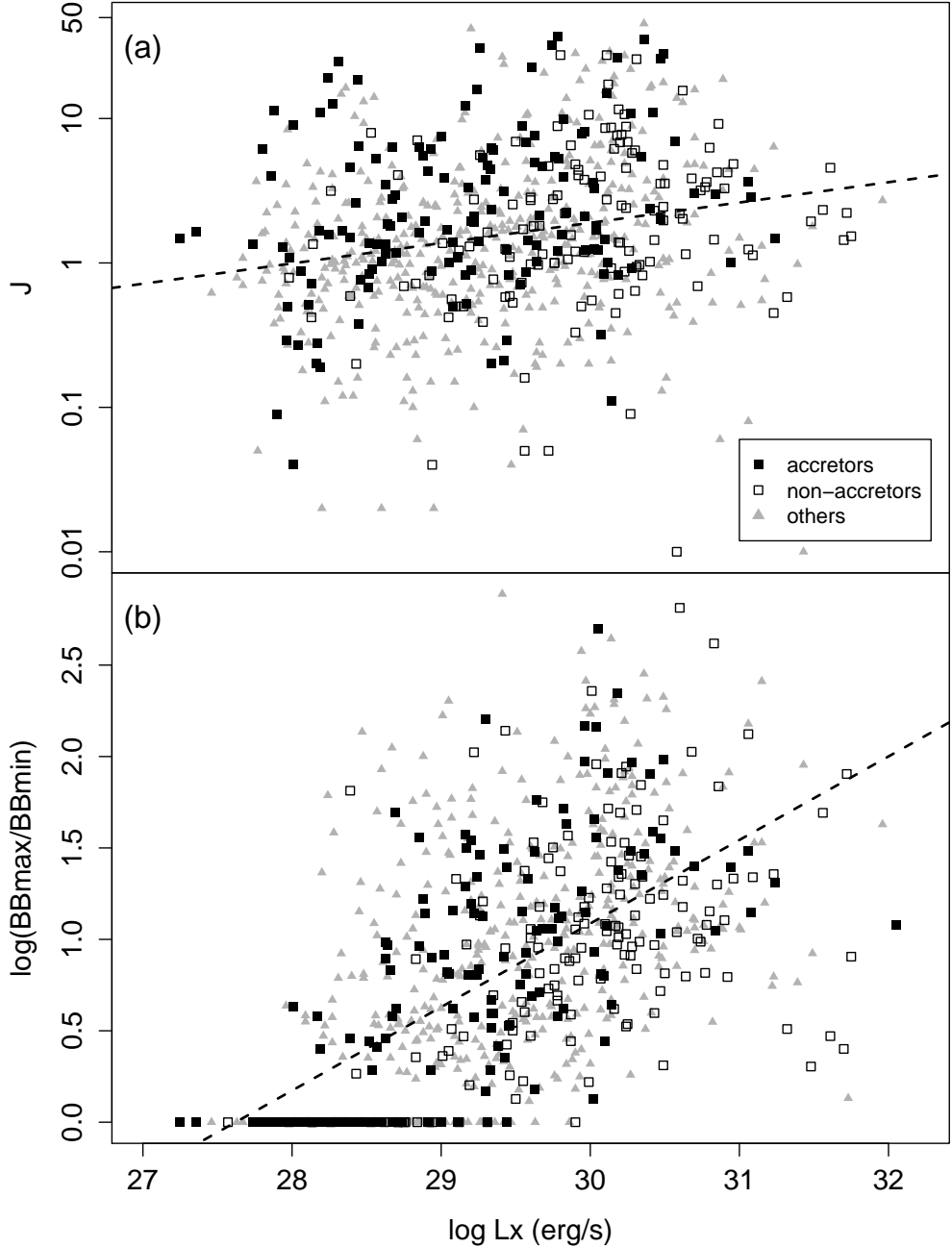}
\caption{\label{var-Lx}
(a) Strength of optical variability as measured by the $J$ statistic is 
plotted vs.\ X-ray luminosity ($L_X$).
(b) Strength of X-ray variability as measured by the BB analysis is
plotted vs.\ $L_X$. Symbols and lines are as in Fig.\ \ref{Jvar-xvar}.
In both panels, a statistically significant correlation in the sense 
of the line is confirmed by a Kendall's $\tau$ test. This correlation 
is in the same sense when the accretors or non-accretors are tested
separately. Only the correlation in (a) is likely to be physical; 
see text.}
\end{figure}

X-ray luminosity ($L_X$) was found by \citet{stass04} to be strongly
correlated with optical variability in their analysis of all previous
{\it Chandra} observations of PMS stars in the ONC. Specifically,
they found that the ONC stars with the highest $L_X$ also exhibited
the highest levels of optical variability. 
This correlation is confirmed here, with a sample that is five times
larger than that studied by \citet{stass04}. The correlation between
$J$ and $L_X$ in our study sample (Fig.\ \ref{var-Lx}a) is highly
statistically significant: The probability that $J$ and $L_X$ are
uncorrelated, according to a Kendall's $\tau$ test, is less than
$10^{-6}$. Importantly, we furthermore find that this correlation
persists, at a less strong but still highly significant level, when
we separately consider accretors and non-accretors. 

At the same time, our sample exhibits a similar correlation between
$L_X$ and X-ray variability (Fig.\ \ref{var-Lx}b), with similarly
high statistical significance as that between $L_X$ and optical
variability. In this case, however, the correlation appears to stem
from an observational bias against detection of X-ray variability
in stars with low $L_X$. Indeed, as shown in Fig.\ \ref{var-Lx}b,
all of the stars for which the BB analysis finds no X-ray variability
(\bbmax/\bbmin\ $=1$) have $L_X < 10^{30}$ erg s$^{-1}$.

That optical and X-ray variability are both so strongly correlated
with $L_X$ suggests that this mutual correlation may be responsible
for the correlation of optical and X-ray variability with another. To
better understand these mutually correlated variables, we perform a
multiple linear least-squares regression analysis on the strength
of optical variability ($J$) simultaneously against the strength
of X-ray variability (\bbmax/\bbmin) and $L_X$. The regression
coefficient for $L_X$ is found to be statistically significant (null
hypothesis probability of $5\times 10^{-6}$), but the coefficient for
\bbmax/\bbmin\ is not (null hypothesis probability of 0.34). Similarly
regressing \bbmax/\bbmin\ simultaneously against $J$ and $L_X$,
we find that the $L_X$ coefficient is statistically significant
(null hypothesis probability of less than $10^{-6}$) but that the
$J$ coefficient is not (null hypothesis probability of 0.34). The
regressions of $J$ and \bbmax/\bbmin\ vs.\ $L_X$ are represented by
dashed lines in Fig.\ \ref{var-Lx}.

Thus, the relationship between optical and X-ray variability in our
sample is, in fact, fully explained by the mutual correlations of $J$
and \bbmax/\bbmin\ with $L_X$. Controlling for this mutual correlation
with $L_X$, we find no statistically significant residual correlation
of optical and X-ray variability with one another. Optical and X-ray
variability are evidently not causally related.

Even though the correlation of optical variability with $L_X$ has
a plausible physical explanation \citep[stars that are more heavily
spotted are also stronger X-ray emitters;][]{stass04}, our finding that
the correlation of X-ray variability with $L_X$ is due to observational
bias (Fig.\ \ref{var-Lx}b) raises the suspicion that the correlation
of optical variability with $L_X$ may be similarly spurious. For
example, $L_X$ is known to be very strongly correlated with bolometric
luminosity ($L_{\rm bol}$) in the COUP sample \citep{preibisch05}. If
optical variability is in turn correlated with $L_{\rm bol}$ in a
manner similar to the correlation of X-ray variability with $L_X$,
then this may explain the correlation of optical variability with $L_X$
as being non-physical.

To examine this possibility quantitatively, we again perform a multiple
regression analysis, as above, but this time on $J$ against $L_{\rm
bol}$ and $L_X$. First of all, we find that optical variability is
indeed correlated with $L_{\rm bol}$, albeit weakly (null-hypothesis
probability of 0.02), when we regress $J$ against $L_{\rm bol}$
alone. However, something unexpected occurs when we include $L_X$
as a regressor. The regression coefficient for $L_{\rm bol}$ becomes
statistically insignificant (null-hypothesis probability of 0.76),
whereas the coefficient for $L_X$ {\it is} significant, and with
a higher degree of significance (null-hypothesis probability
of 0.004\footnote{Note that the statistical significance of the
correlation between $J$ and $L_X$ is lower here than in the multiple
regression analysis of $J$ on $L_X$ and \bbmax/\bbmin. This is because
the sample under consideration here is smaller (545 stars vs.\ 814
stars), since $L_{\rm bol}$ measurements are not available for all
stars in the COUP database.}) than was the correlation of $J$ with
$L_{\rm bol}$ alone. In other words, controlling for the correlation of
$J$ with $L_{\rm bol}$ does {\it not} eliminate the correlation of $J$
with $L_X$; on the contrary, controlling for the correlation of $J$
with $L_X$ eliminates the correlation of $J$ with $L_{\rm bol}$. $L_X$
is evidently the more fundamental correlate.

This somewhat surprising result reinforces the conclusion that optical
variability is intrinsically related to X-ray luminosity in our study
sample. Having considered in detail the mutual correlations of $J$ with
\bbmax/\bbmin, $L_{\rm bol}$, and $L_X$, only the correlation between
$J$ and $L_X$ remains. The correlation between optical variability
and $L_X$ is more significant even than the intuitive correlation of
optical variability with accretion (\S \ref{var-accretion}). Indeed,
after controlling for the influence of all other secondary variables,
we find that {\it the strongest correlate of optical variability in
our study sample is X-ray luminosity}.

\subsection{Summary\label{summary}}
We find that optical and X-ray variability in our study sample are
not causally related. This conclusion follows from an examination of
the joint occurrences of time-averaged optical and X-ray variability,
as well as their mutual correlations with one another and with other
variables. We find definitive evidence that, on timescales of 7~d
or less, optical variability does not always occur in the presence
of X-ray variability, nor vice-versa. Even when the two forms of
variability are present, non-accreting stars are just as likely to show
both forms of variability as are stars that exhibit active accretion;
that is, among stars exhibiting both optical and X-ray variability,
accretion is evidently not the cause. At a more detailed level,
we have performed a multiple regression analysis, controlling for
the mutual correlations of optical and X-ray variability, both with
one another and with bolometric and X-ray luminosity. This analysis
reveals that, while statistically significant correlations exist
among all of these measures, the fundamental correlation in this
study is that between optical variability and X-ray luminosity. This
correlation is furthermore independent of accretion, indicating again
that accretion is not the cause.

\section{Discussion\label{discussion}}
While magnetic activity, and therefore X-ray production, in late-type
main-sequence stars is relatively well understood in terms of
a rotation-driven dynamo \citep[e.g.][]{schrijver00}, whether
stellar rotation is also at the heart of PMS X-ray production is
a contested issue. In contrast to the strong ``rotation-activity
relationship" that is clearly present on the main sequence
\citep{pallavicini81,jeffries99,randich00,pizzolato03}, a series
of studies based on {\it Chandra} observations of a large sample
of PMS stars in the Orion Nebula Cluster (ONC) failed to find any
direct correlation between the stars' rotation rates and their X-ray
luminosities \citep{flac-time,feig03,preibisch05}.

However, \citet{stass04}, in an analysis of all archival {\it Chandra}
observations of PMS stars in the ONC, found indirect evidence hinting
that a main--sequence-type rotation-activity relationship may in fact
be driving PMS X-ray production after all. They found a statistically
significant, positive correlation between stellar rotation period
and $L_X$, suggesting that these stars are on the ``super-saturated"
portion of the rotation-activity relation, consistent with their Rossby
numbers. They furthermore found that the sample of ONC stars with
known rotation periods is significantly biased to high $L_X$; stars
whose rotation periods are unknown, and which have lower $L_X$, may
thus represent the unseen ``linear" portion of the rotation-activity
relation. \citet{preibisch05} and \citet{rebull06} have since confirmed
these findings in different Orion samples.

The results of the present study lend further support for an
interpretation in which accretion is {\it not} a significant source
of X-ray production in PMS stars. Accretion, as traced by emission in
\ion{Ca}{2}, does not manifest itself in the X-ray variability of our
study sample (\S \ref{var-accretion}). And while accretion {\it is}
manifested in the {\it optical} variability of these stars as expected,
a multiple regression analysis finds that X-ray variability is not
correlated with this optical variability (\S \ref{var-lx}). Moreover,
we have found very few cases in our study sample in which the optical
and X-ray variability are correlated in time (Paper~I).

That we find so little evidence for accretion as a dominant
source of X-rays in young stars suggests that we look to other
possible X-ray production mechanisms. Indeed, there are now strong
indications that X-ray production is intimately related in some
way to the presence of magnetically active surface regions on these
stars. The evidence for the existence of magnetically active surface
regions on PMS stars is abundant. Several important studies of
Zeeman broadening in the photospheric spectral lines of PMS stars
\citep[e.g.][]{johnskrull99,guenther99,johnskrull04}, as well as
Doppler imaging of their surfaces \citep[e.g.][]{rice96,johnskrull97},
provide direct proof of the existence of strong surface
fields. Moreover, the X-ray flares observed in the COUP study can only
be reasonably explained as reconnection events in strong stellar fields
\citep{feig06}. In addition, numerous photometric studies of PMS stars
have relied upon the presence of surface features to measure these
stars' rotation periods \citep[e.g.][]{stass99,rebull01,herbst02}. The
stability of these features over timescales of up to several years
strongly suggests that these features are in many cases ``starspots"
presumably analogous to dark spots on the Sun.

\citet{stass04} found that PMS stars in the ONC exhibit a highly
statistically significant correlation between the strength of
their optical variability (taken to be a measure of the stars'
``spottedness") and their X-ray luminosities, in the sense that
the least X-ray luminous stars are those with the smallest amount
of spottedness. This finding led those authors to infer a causal
connection between X-ray emission and the presence of starspots,
leading to the conclusion that solar-analog surface magnetic activity
driving hot coronae is the primary source of X-ray production in PMS
stars. However, the possibility remained that the correlation between
optical variability and X-ray luminosity was driven by accretion.

The results of the present study firmly corroborate the correlation
between optical variability and X-ray luminosity (i.e.\ Fig.\
\ref{var-Lx})---indeed, optical variability is the strongest correlate
of X-ray luminosity in the COUP study sample (\S \ref{var-lx})---and
furthermore show that this correlation exists regardless of whether
the stars under consideration are accreting or non-accreting (\S
\ref{var-accretion}). This finding contradicts models in which the
observed X-ray emission is produced primarily by accretion, and instead
bolsters support for the role of non--accretion-related, optically
active magnetic regions (i.e.\ spots) in the production of PMS X-rays.

A small number of the stars in our sample evince periodic X-ray
variability with a periodicity that is very similar to the observed
periods in the optical \citep{flac-spots}. In these cases the sites of
X-ray emission may be predominantly co-spatial with magnetic surface
spots. However, the association of X-ray production with magnetic spots
does not manifest itself in a time-correlated way for most stars in
our sample (Paper~I), indicating that the X-ray emitting coronae in PMS
stars generally have spatial structures that little reflect the spatial
distribution of magnetic footpoints in the underlying photosphere.

These conclusions are consistent with recent studies of TTS magnetic
field structures. For example, \citet{jardine06} find that a variety
of field topologies are required to explain the observed scatter of
emission measures of COUP stars, ranging from simple dipole fields to
much more complex configurations. As discussed by those authors, the
more complex field configurations are characterized by coronal gas that
is confined to compact loops covering a large fraction of the stellar
surface. Coronal emission arising from such a topology will have a
relatively low X-ray luminosity by virtue of its compact size, and
the footpoints of the multi-polar field will produce correspondingly
small spots that are more-or-less uniformly distributed on the
stellar surface \citep[cf.\ Fig.\ 8 in][]{jardine06}, resulting in a
relatively low-amplitude spot signal in the optical. The reverse will
be true for a star whose coronal emission arises in a more extended,
more dipolar field, rooted in a small number of footpoints (spots)
that are more asymmetrically distributed on the stellar surface; such
a configuration will produce a high X-ray luminosity together with
strong optical variability. Thus, given a large ensemble of PMS stars
with a range of field configurations, a strong correlation between
X-ray luminosity and optical ``spottedness," as we have found here,
may naturally follow.

\section{Conclusions\label{conclusions}}
We have presented a combined analysis of time-averaged optical and
X-ray variability for more than 800 pre--main-sequence (PMS) stars
in the Orion Nebula Cluster (ONC) with the aim of elucidating the
origins of X-ray production in PMS stars. This study complements
and extends the analysis of time-correlated variability in this same
sample reported in \citet{stass06}.

Combining the results of these two studies, we find very little
evidence to suggest a direct physical link between the sources of
optical and X-ray variability in these stars. Whether we consider
time-correlated behavior between the individual stars' simultaneous
optical and X-ray light curves, whether we consider the optical and
X-ray variability properties of our study sample as a statistical
ensemble, whether we segregate the stars according to their accretion
properties, we arrive at the same conclusion: the sites of optical
and X-ray variability in PMS stars are not---for the vast majority of
stars---instantaneously one and the same. An interpretation in which
accretion is a dominant source of X-ray production in PMS stars is
not supported by the data.

Instead, the data provide strong support for an interpretation in
which X-ray production in PMS stars is intimately connected with
the presence and strength of optically variable, magnetic regions
on their surfaces (i.e., spots). Indeed, after controlling for the
influence of all other secondary variables, including accretion, we
find that the strongest correlate of X-ray luminosity in our sample
is spottedness (as measured by optical variability, after controlling
for accretion). Our interpretation is that solar-analog magnetic
activity is the primary, common driving mechanism of both optical
variability and X-ray production in these stars. The fact that this
relationship between spottedness and X-ray production does not, in
the vast majority of cases, manifest itself in a time-correlated way
(Paper~I) suggests that the optical spots represent the footpoints
of complex magnetic topologies that heat solar-analog coronae, as
suggested by \citet{jardine06}.

\acknowledgments
It is a pleasure to acknowledge the {\it Chandra} Orion Ultradeep
Project (COUP) team, supported by {\it Chandra} Guest Observer grant
SAO GO3--4009A (PI: E.\ Feigelson). This work is also supported by
an NSF Career award (AST--0349075), and by a Cottrell Scholar award
from the Research Corporation, to K.~G.~S.

Facility: CXO(ACIS)

\clearpage

\begin{deluxetable}{rrrrrr}
\tablecolumns{6}
\tablewidth{0pt}
\tabletypesize{\scriptsize}
\tablecaption{Optical and X-ray Variability in Study Sample\label{var-table}}
\tablehead{
\colhead{COUP} & \colhead{Opt.\tablenotemark{a}} & 
\colhead{$J$\tablenotemark{b}} &
\colhead{$\log P_{\rm KS}$\tablenotemark{c}} & \colhead{BB\tablenotemark{d}} & 
\colhead{\ion{Ca}{2}\tablenotemark{e}}
}
\startdata
   6 & 40 &  1.65 & $\le -$4.00 &  11.5 &   0.0 \\
   7 & 45 &  0.51 & $\le -$4.00 &  11.3 &  \nodata \\
   9 & 46 &  1.21 & $\le -$4.00 &  81.8 &  \nodata \\
  10 & 48 &  0.59 & $-$3.00 &  65.0 &   1.0 \\
  11 & 50 & 35.44 & $\le -$4.00 &  29.5 & $-$14.6 \\
  12 & 51 & 10.80 & $\le -$4.00 &   7.2 &   0.0 \\
  13 & 53 &  4.06 & $-$1.01 &   1.0 &   1.5 \\
  14 & 54 &  0.85 & $-$2.70 &   3.2 &   0.0 \\
  15 & 55 &  8.85 & $\le -$4.00 &   4.6 &   1.0 \\
  16 & 3062 &  1.03 & $-$0.26 &   1.0 &  $-$3.2 \\
  17 & 63 & 10.62 & $\le -$4.00 &  16.9 &   1.9 \\
  20 & 70 &  0.29 & $-$2.85 &   4.1 &   0.0 \\
  21 & 71 &  0.45 & $\le -$4.00 &   3.3 &   0.0 \\
  23 & 75 &  0.77 & $\le -$4.00 &  15.8 &  \nodata \\
  26 & 10251 &  1.61 & $-$1.17 &   1.0 &  \nodata \\
  27 & 77 & 11.57 & $\le -$4.00 &  11.6 &   1.8 \\
  28 & 81 &  9.18 & $\le -$4.00 &  68.6 &   1.6 \\
  29 & 83 &  9.92 & $\le -$4.00 &   4.2 &  $-$9.2 \\
  30 & 84 &  1.15 & $-$0.47 &   1.0 &   0.0 \\
  33 & 10259 & $-$0.52 & $-$0.96 &   1.0 &  \nodata \\
  37 & 90 & $-$0.13 & $-$1.70 &   1.0 &   0.7 \\
  40 & 92 &  0.60 & $\le -$4.00 &  52.9 &   0.0 \\
  41 & 98 & $-$0.53 & $\le -$4.00 &  67.1 &   0.0 \\
  43 & 99 & 25.76 & $\le -$4.00 &  51.0 &   1.4 \\
  44 & 104 &  8.99 & $\le -$4.00 &   4.3 & $-$10.7 \\
  45 & 102 &  9.27 & $-$0.56 &   1.0 &   0.0 \\
  46 & 105 & $-$0.46 & $-$1.60 &   1.0 &   0.0 \\
  47 & 106 &  1.10 & $\le -$4.00 &   3.4 &   1.5 \\
  49 & 107 &  6.22 & $\le -$4.00 &   3.3 &  $-$9.6 \\
  50 & 110 &  0.02 & $-$1.00 &   1.0 &   0.0 \\
  54 & 113 & 22.49 & $\le -$4.00 &   4.9 &  $-$1.0 \\
  55 & 114 & 15.89 & $\le -$4.00 &   6.4 &  $-$1.4 \\
  57 & 116 &  1.44 & $\le -$4.00 &   4.0 &   1.6 \\
  58 & 117 &  6.34 & $-$2.60 &   9.2 &  $-$6.3 \\
  59 & 10281 &  1.15 & $\le -$4.00 &   2.1 &  \nodata \\
  60 & 120 &  1.30 & $-$2.49 &   1.6 &   1.0 \\
  62 & 123 & 16.29 & $\le -$4.00 &  26.8 &   0.0 \\
  64 & 124 & $-$0.56 & $\le -$4.00 &   8.2 &   0.8 \\
  65 & 125 &  7.36 & $\le -$4.00 &  17.0 &   0.4 \\
  66 & 127 & 10.92 & $\le -$4.00 &  30.4 &  $-$2.8 \\
  67 & 128 &  1.24 & $\le -$4.00 &  13.0 &   0.0 \\
  68 & 130 & $-$0.47 & $-$2.08 &   3.5 &  \nodata \\
  69 & 132 &  1.63 & $\le -$4.00 &   2.1 &  \nodata \\
  71 & 133 &  2.40 & $\le -$4.00 &  10.7 &   1.6 \\
  72 & 135 &  6.12 & $\le -$4.00 &   1.9 &  $-$2.3 \\
  73 & 134 &  1.47 & $-$1.83 &   3.4 &  \nodata \\
  74 & 137 &  1.85 & $\le -$4.00 &   3.2 &  \nodata \\
  75 & 136 & 21.94 & $-$3.52 &   1.8 &  \nodata \\
  77 & 138 &  6.31 & $-$1.01 &   1.0 &  $-$2.8 \\
  80 & 10294 &  0.35 & $-$0.61 &   1.0 &  \nodata \\
  85 & 143 &  0.20 & $-$1.95 &   1.0 &  $-$2.3 \\
  86 & 144 & $-$0.39 & $-$1.42 &   1.0 &   0.0 \\
  88 & 145 & $-$0.15 & $\le -$4.00 & 183.3 &  \nodata \\
  89 & 146 &  1.77 & $\le -$4.00 &  28.7 &   0.0 \\
  90 & 148 &  1.45 & $\le -$4.00 & 415.2 &   1.6 \\
  94 & 149 &  1.50 & $-$1.14 &   2.9 & $-$23.7 \\
  95 & 151 & $-$0.11 & $-$0.29 &   1.0 &  $-$5.3 \\
  96 & 152 &  1.80 & $\le -$4.00 &  33.9 &   2.6 \\
  97 & 3111 &  1.96 & $\le -$4.00 &  13.8 &  $-$3.4 \\
 100 & 153 &  0.70 & $\le -$4.00 &  11.1 &  \nodata \\
 101 & 155 &  2.20 & $\le -$4.00 & 650.4 &   2.9 \\
 102 & 3046 &  0.76 & $-$1.61 &   1.0 &   0.0 \\
 107 & 157 &  0.98 & $\le -$4.00 &   5.8 &  \nodata \\
 108 & 158 &  1.02 & $\le -$4.00 &  16.7 &   1.5 \\
 109 & 159 &  1.77 & $\le -$4.00 &   3.9 &   0.3 \\
 110 & 10308 &  0.45 & $\le -$4.00 &  20.5 &  \nodata \\
 112 & 164 &  5.65 & $\le -$4.00 &  69.5 &  $-$0.7 \\
 113 & 165 & 15.86 & $\le -$4.00 &  11.6 &  \nodata \\
 114 & 168 &  0.66 & $-$2.42 &   4.1 &   0.0 \\
 115 & 167 &  0.09 & $\le -$4.00 &   8.2 &   1.4 \\
 117 & 169 &  6.62 & $\le -$4.00 &  32.2 &  \nodata \\
 118 & 171 &  0.44 & $\le -$4.00 &   4.3 &   0.4 \\
 119 & 174 &  1.63 & $-$1.85 &   1.0 &   3.7 \\
 122 & 175 & 19.54 & $\le -$4.00 & 204.8 &   0.0 \\
 123 & 176 &  4.45 & $\le -$4.00 &  33.7 &   0.8 \\
 124 & 176 &  4.45 & $\le -$4.00 &  18.1 &   0.8 \\
 125 & 181 &  0.23 & $\le -$4.00 &   2.8 &   0.8 \\
 126 & 177 &  0.77 & $\le -$4.00 &  38.8 &   0.9 \\
 127 & 182 & $-$0.12 & $-$0.17 &   1.0 &  \nodata \\
 128 & 183 &  1.46 & $-$0.48 &   1.0 &   0.0 \\
 130 & 184 &  0.56 & $\le -$4.00 &   3.2 &   1.6 \\
 131 & 187 &  2.45 & $\le -$4.00 &  44.8 &   1.4 \\
 132 & 188 &  1.70 & $\le -$4.00 &  33.6 &   0.0 \\
 133 & 189 &  0.49 & $-$0.18 &   1.0 &   0.0 \\
 134 & 190 & $-$0.09 & $-$3.70 &   7.8 &   2.0 \\
 137 & 191 &  3.14 & $\le -$4.00 &   6.6 &   0.0 \\
 139 & 192 & 17.46 & $\le -$4.00 &  12.7 &   0.9 \\
 141 & 193 &  6.90 & $\le -$4.00 &  30.4 & $-$17.8 \\
 142 & 197 &  0.79 & $\le -$4.00 &  64.4 &  \nodata \\
 143 & 196 &  2.84 & $\le -$4.00 &  11.4 &   1.2 \\
 144 & 5106 &  5.95 & $\le -$4.00 &   7.5 &   0.0 \\
 147 & 198 &  5.43 & $\le -$4.00 &  15.0 &  $-$1.0 \\
 148 & 200 &  1.08 & $-$0.26 &   1.0 &   0.0 \\
 149 & 202 &  2.59 & $-$1.34 &   1.0 &  $-$2.1 \\
 150 & 201 &  0.61 & $\le -$4.00 &  21.9 &   1.5 \\
 152 & 206 &  3.47 & $\le -$4.00 &  14.5 &  \nodata \\
 153 & 209 &  1.65 & $-$0.71 &   1.0 &  \nodata \\
 154 & 205 &  0.60 & $\le -$4.00 &   6.0 &   0.0 \\
 157 & 207 &  1.15 & $-$0.45 &   1.0 &  \nodata \\
 158 & 208 &  3.03 & $\le -$4.00 &   4.2 &   0.0 \\
 159 & 210 &  4.88 & $\le -$4.00 & 156.2 &  $-$0.5 \\
 161 & 211 &  4.68 & $\le -$4.00 &   5.4 &   1.4 \\
 164 & 213 & $-$0.34 & $-$0.73 &   1.0 &   0.6 \\
 166 & 212 &  0.23 & $\le -$4.00 &   2.2 &   0.0 \\
 168 & 215 &  0.15 & $-$3.40 &   6.1 &  \nodata \\
 169 & 218 &  1.74 & $\le -$4.00 &   2.1 &   0.0 \\
 171 & 217 & $-$0.02 & $-$0.11 &   1.0 &   0.0 \\
 173 & 220 &  4.77 & $\le -$4.00 &  17.5 &   1.0 \\
 174 & 222 &  4.04 & $\le -$4.00 &  13.2 &   1.2 \\
 176 & 3138 &  0.97 & $-$0.94 &   1.9 &  \nodata \\
 177 & 223a &  2.75 & $-$2.96 &  19.0 &   2.0 \\
 180 & 224 &  4.58 & $-$1.05 &   1.0 &  $-$4.9 \\
 181 & 227 &  0.75 & $-$0.49 &   1.0 &  \nodata \\
 183 & 226 &  0.96 & $\le -$4.00 &  17.6 &   0.0 \\
 187 & 230 &  1.05 & $-$0.71 &   1.0 &   0.0 \\
 188 & 232 & 18.72 & $\le -$4.00 &  15.5 &  \nodata \\
 189 & 229 &  6.92 & $\le -$4.00 &   1.3 &   2.1 \\
 190 & 231 &  4.87 & $-$0.52 &   1.0 &   0.0 \\
 192 & 5042 &  0.54 & $\le -$4.00 &  14.9 &  \nodata \\
 193 & 233 &  1.67 & $-$0.63 &   1.0 &  $-$4.9 \\
 194 & 234 & $-$0.39 & $-$0.03 &   1.0 &   0.0 \\
 197 & 235 & 26.74 & $\le -$4.00 &   4.3 &  \nodata \\
 199 & 5100 &  2.96 & $\le -$4.00 &  49.2 & $-$18.3 \\
 200 & 236 &  3.49 & $-$0.71 &   1.0 &  \nodata \\
 201 & 237 &  6.23 & $\le -$4.00 &   3.9 &   0.0 \\
 202 & 239 &  8.59 & $\le -$4.00 &  12.1 &   1.5 \\
 205 & 243 &  0.82 & $\le -$4.00 &  22.0 &   1.2 \\
 206 & 10362 &  0.75 & $-$0.17 &   1.0 &  \nodata \\
 207 & 241 &  1.10 & $-$0.05 &   1.0 &  \nodata \\
 208 & 244 &  0.19 & $-$3.52 &   2.5 &  $-$1.7 \\
 212 & 245 &  5.27 & $-$1.23 &   1.0 &  $-$5.8 \\
 214 & 248 & 20.68 & $\le -$4.00 &  15.6 &  $-$0.9 \\
 217 & 249 & 27.50 & $\le -$4.00 &  23.6 &   1.7 \\
 218 & 250 &  2.36 & $\le -$4.00 &   9.6 &   0.9 \\
 219 & 254 &  3.15 & $-$1.71 &   1.0 &   5.6 \\
 221 & 251 &  0.72 & $-$3.70 &   2.3 &   2.0 \\
 222 & 256 &  0.50 & $\le -$4.00 &   9.0 &   2.1 \\
 223 & 253 & 15.62 & $\le -$4.00 &  20.9 &   1.7 \\
 226 & 258 &  8.79 & $\le -$4.00 &  88.0 &   1.2 \\
 227 & 259 &  4.93 & $-$1.87 &   1.5 &  $-$2.7 \\
 228 & 257 &  0.12 & $-$1.53 &   1.0 &   0.0 \\
 230 & 10368 &  0.58 & $\le -$4.00 &   6.6 &  \nodata \\
 232 & 261 &  8.39 & $-$2.05 &   1.0 &  \nodata \\
 234 & 5112 &  0.97 & $-$1.05 &   1.0 &  \nodata \\
 235 & 262 & $-$0.12 & $-$0.19 &   1.0 &   0.0 \\
 236 & 265 &  3.80 & $\le -$4.00 &   3.9 &   0.0 \\
 237 & 5096 &  1.31 & $\le -$4.00 &   1.9 &   0.0 \\
 238 & 266 & 10.99 & $\le -$4.00 &  14.3 &  \nodata \\
 240 & 270 & $-$0.19 & $-$1.34 &   2.7 &   0.0 \\
 241 & 268 &  5.58 & $\le -$4.00 &  13.5 &   2.8 \\
 242 & 10370 &  1.52 & $\le -$4.00 &  38.3 &  \nodata \\
 243 & 272 &  3.49 & $-$1.83 &   7.8 &  $-$2.2 \\
 244 & 3040 & $-$0.44 & $\le -$4.00 &   9.5 &  \nodata \\
 245 & 273 &  3.65 & $\le -$4.00 &  30.4 & $-$15.9 \\
 246 & 277 &  2.07 & $-$2.27 &   1.0 &   0.0 \\
 248 & 279 &  1.65 & $-$0.08 &   1.0 &  $-$3.0 \\
 249 & 276 &  1.15 & $\le -$4.00 &  56.2 &   1.1 \\
 250 & 278 & 30.56 & $\le -$4.00 &  28.9 &  $-$9.6 \\
 252 & 275a &  3.77 & $\le -$4.00 &  15.1 &   1.8 \\
 253 & 280 &  0.55 & $\le -$4.00 & 228.3 &   1.4 \\
 255 & 281 &  0.77 & $\le -$4.00 &   4.9 &   1.3 \\
 256 & 283 &  0.26 & $\le -$4.00 &   2.7 &   0.0 \\
 259 & 285 &  1.29 & $\le -$4.00 &   6.2 &  $-$0.4 \\
 260 & 3064 &  1.14 & $\le -$4.00 &  14.1 &  \nodata \\
 262 & 286 &  0.69 & $\le -$4.00 &  10.1 &   2.3 \\
 264 & 288 &  1.03 & $-$1.14 &   1.0 &   0.9 \\
 265 & 287 &  0.09 & $-$1.92 &   1.0 &  $-$4.3 \\
 266 & 290 &  0.82 & $\le -$4.00 &   2.6 &  \nodata \\
 270 & 291 &  1.38 & $\le -$4.00 &  17.6 &   1.5 \\
 271 & 292 &  0.80 & $-$2.41 &   3.3 &   0.0 \\
 273 & 5159 &  1.77 & $-$0.19 &   1.0 &  \nodata \\
 275 & 296 &  0.38 & $-$1.08 &   1.0 &  $-$3.2 \\
 276 & 298 &  0.66 & $\le -$4.00 &  58.5 &  \nodata \\
 279 & 299 &  0.92 & $-$0.16 &   1.0 &  \nodata \\
 281 & 5064 &  0.44 & $-$0.18 &   1.0 &   0.0 \\
 283 & 300 &  1.20 & $\le -$4.00 & 185.8 &   0.0 \\
 286 & 302 &  0.73 & $-$2.54 &   6.5 &  \nodata \\
 287 & 5138 &  0.70 & $-$0.25 &   1.0 &  \nodata \\
 289 & 303 &  0.55 & $-$0.54 &   1.0 &  \nodata \\
 290 & 304 &  0.87 & $\le -$4.00 & 376.8 &   0.6 \\
 291 & 5075 &  5.74 & $-$1.36 &   1.0 &  \nodata \\
 294 & 306 &  1.40 & $-$0.46 &   2.1 &   0.0 \\
 296 & 307 &  0.82 & $-$2.13 &   1.0 &   1.7 \\
 298 & 311 &  3.63 & $\le -$4.00 &   3.0 &   0.0 \\
 300 & 309 &  3.05 & $-$1.37 &   1.0 &   0.8 \\
 301 & 313 & 22.07 & $\le -$4.00 & 259.3 &   0.5 \\
 304 & 314 &  1.69 & $\le -$4.00 &  38.1 &  \nodata \\
 305 & 318 &  0.59 & $-$1.32 &   1.0 &   0.8 \\
 309 & 317 &  2.36 & $\le -$4.00 &  11.3 &  \nodata \\
 310 & 319 &  0.64 & $\le -$4.00 &  13.5 &   8.9 \\
 314 & 320 &  7.36 & $\le -$4.00 &  23.4 &   0.3 \\
 316 & 322 &  1.35 & $\le -$4.00 &   2.9 &  $-$1.8 \\
 318 & 323 & 14.06 & $-$3.30 &  12.2 &   0.8 \\
 320 & 10395 &  0.46 & $-$0.55 &   1.0 &  \nodata \\
 321 & 324 &  0.89 & $\le -$4.00 &  34.8 &  $-$3.3 \\
 322 & 3072 &  1.15 & $\le -$4.00 &  31.3 &  \nodata \\
 323 & 326 &  1.21 & $\le -$4.00 &  28.8 &   2.0 \\
 325 & 328 & 29.11 & $\le -$4.00 &   9.4 &   0.0 \\
 326 & 3045 &  3.67 & $-$1.31 &   1.0 &  \nodata \\
 328 & 330 &  1.97 & $\le -$4.00 &   2.0 &   1.8 \\
 331 & 3134 &  0.16 & $\le -$4.00 & 181.2 &  \nodata \\
 333 & 331 &  0.75 & $\le -$4.00 &  50.0 &  \nodata \\
 336 & 332 &  1.11 & $-$2.60 &   1.0 &  $-$0.9 \\
 339 & 335 &  0.88 & $\le -$4.00 &   7.9 & $-$26.0 \\
 340 & 333 &  0.50 & $\le -$4.00 &   2.9 &   3.7 \\
 341 & 334 &  7.49 & $-$0.76 &   1.0 &  $-$4.8 \\
 342 & 336 &  0.85 & $\le -$4.00 &   8.9 &  \nodata \\
 343 & 337 &  4.84 & $\le -$4.00 &  21.5 &   1.9 \\
 346 & 338 &  2.42 & $\le -$4.00 &  12.7 &   0.0 \\
 349 & 1772 &  1.03 & $-$1.82 &   1.0 &  \nodata \\
 350 & 9002 &  0.96 & $-$1.06 &   1.0 &  \nodata \\
 358 & 340 &  0.10 & $\le -$4.00 &  57.0 &  \nodata \\
 362 & 339 &  3.97 & $\le -$4.00 &  26.1 &   0.0 \\
 363 & 342 &  3.35 & $-$0.55 &   1.0 &  \nodata \\
 365 & 341 &  0.62 & $\le -$4.00 &  21.3 &   0.0 \\
 367 & 3047 &  1.19 & $\le -$4.00 &   6.3 &   0.0 \\
 368 & 343 &  3.65 & $\le -$4.00 &   3.0 &   0.9 \\
 378 & 345 &  2.98 & $\le -$4.00 &   3.7 &   0.0 \\
 379 & 3104 &  1.00 & $\le -$4.00 &   2.8 &  \nodata \\
 380 & 5142 &  0.92 & $-$1.18 &   1.0 &  \nodata \\
 382 & 347 &  0.49 & $\le -$4.00 &  23.0 &   0.0 \\
 383 & \nodata &  1.73 & $-$1.39 &   1.0 &  \nodata \\
 385 & 349 &  1.95 & $\le -$4.00 &  32.1 &   0.0 \\
 387 & 348 &  7.37 & $\le -$4.00 &   8.7 &  \nodata \\
 394 & 352 &  0.58 & $\le -$4.00 &   3.2 &   1.9 \\
 395 & 354 &  0.50 & $\le -$4.00 &  21.4 &   1.3 \\
 399 & 3102 & $-$0.05 & $\le -$4.00 & 201.5 &  \nodata \\
 403 & 355 &  2.23 & $\le -$4.00 &  43.0 &  $-$6.2 \\
 404 & 356 &  1.57 & $\le -$4.00 &  13.4 &  $-$1.3 \\
 405 & 357 &  0.80 & $-$3.10 &   4.0 &   0.0 \\
 408 & 10415 & $-$0.18 & $-$1.25 &   1.0 &  \nodata \\
 410 & 359 &  0.30 & $\le -$4.00 &  18.3 &  \nodata \\
 411 & 3098 &  0.59 & $-$1.54 &   1.0 &  \nodata \\
 414 & 358 &  1.95 & $\le -$4.00 &  97.1 &   0.0 \\
 416 & \nodata &  4.28 & $-$2.10 &   1.0 &  \nodata \\
 417 & 361 &  4.10 & $\le -$4.00 &  25.2 &  \nodata \\
 421 & 3005 & $-$0.42 & $-$0.18 &   1.0 &   0.0 \\
 430 & 365 &  1.31 & $\le -$4.00 &  33.8 &  \nodata \\
 431 & 364 & $-$0.14 & $\le -$4.00 &  12.0 &  \nodata \\
 433 & 3070 &  2.51 & $-$0.17 &   1.0 &  \nodata \\
 434 & 366 &  0.27 & $-$0.55 &   1.0 &  $-$2.7 \\
 437 & 371 &  0.26 & $-$1.15 &   1.0 &  \nodata \\
 439 & 369 &  3.68 & $\le -$4.00 &   2.3 &  \nodata \\
 442 & 9008 &  0.86 & $\le -$4.00 &   8.1 &   0.0 \\
 443 & 368 &  2.65 & $-$0.51 &   1.0 &  \nodata \\
 446 & 367 &  0.23 & $\le -$4.00 &   3.8 &  \nodata \\
 449 & 3058 &  0.43 & $\le -$4.00 &  26.5 &   0.0 \\
 450 & \nodata &  2.71 & $\le -$4.00 &  42.5 &  \nodata \\
 452 & 370 &  0.94 & $\le -$4.00 &   6.9 &   1.4 \\
 454 & 373 &  6.28 & $\le -$4.00 &  14.2 &   2.1 \\
 459 & 375 &  1.89 & $\le -$4.00 &  38.3 &  \nodata \\
 462 & 5132 &  0.86 & $-$0.02 &   1.0 &  \nodata \\
 467 & 379 &  2.06 & $\le -$4.00 &  10.5 &  \nodata \\
 468 & 380 & $-$0.45 & $\le -$4.00 &  74.7 &  \nodata \\
 470 & 378a &  3.55 & $\le -$4.00 &   6.5 &   1.5 \\
 474 & 381 & 41.97 & $-$2.96 &   7.1 &   0.0 \\
 478 & \nodata &  2.28 & $-$3.10 &   2.0 &  \nodata \\
 480 & \nodata &  4.28 & $-$0.76 &   1.0 &  \nodata \\
 482 & 5074 & $-$0.01 & $-$3.52 &   3.1 &  \nodata \\
 485 & 387 &  0.92 & $\le -$4.00 &  93.2 &  $-$1.0 \\
 486 & 382 &  0.20 & $\le -$4.00 &   5.0 &   0.0 \\
 487 & \nodata &  1.52 & $\le -$4.00 & 167.7 &  \nodata \\
 488 & 385 & $-$0.76 & $\le -$4.00 &   3.7 &  \nodata \\
 489 & 383 &  3.70 & $-$2.31 &   4.3 &   0.0 \\
 490 & 386 & $-$0.35 & $\le -$4.00 &  69.9 &   1.2 \\
 492 & 389 &  2.54 & $\le -$4.00 &   3.5 &   4.0 \\
 493 & 390 &  1.12 & $\le -$4.00 &   2.0 &  \nodata \\
 497 & \nodata &  1.47 & $-$2.59 &   1.0 &  \nodata \\
 498 & 392 & $-$0.17 & $-$1.09 &   1.0 &   0.0 \\
 499 & 388 &  7.64 & $\le -$4.00 &   9.4 &   1.2 \\
 501 & 391 &  8.82 & $\le -$4.00 &  14.3 & $-$25.2 \\
 513 & 397 & $-$0.08 & $-$0.16 &   1.0 &  \nodata \\
 514 & 393 & $-$0.75 & $\le -$4.00 &   4.8 &  \nodata \\
 515 & 394 &  3.14 & $\le -$4.00 &  16.4 &   0.0 \\
 516 & 9028 &  0.29 & $\le -$4.00 &  24.8 &  $-$2.8 \\
 517 & 395 &  1.40 & $\le -$4.00 &   4.2 &  $-$6.2 \\
 518 & 9029 &  0.53 & $\le -$4.00 &   3.2 &   1.5 \\
 522 & 9032 &  2.09 & $\le -$4.00 &  13.9 & $-$30.4 \\
 527 & 405 &  1.66 & $\le -$4.00 &   6.3 &  \nodata \\
 528 & 400 &  0.20 & $\le -$4.00 &  74.2 &   0.0 \\
 533 & \nodata &  1.83 & $\le -$4.00 & 774.0 &  \nodata \\
 534 & 401 &  1.23 & $-$0.82 &   1.0 & $-$22.3 \\
 535 & 405 &  1.66 & $-$2.06 &   3.5 &  \nodata \\
 536 & 403 &  1.20 & $\le -$4.00 &  11.3 &   1.6 \\
 538 & 404 &  0.91 & $-$0.87 &   1.0 &  \nodata \\
 543 & 406 &  2.94 & $\le -$4.00 &   4.9 &   1.7 \\
 545 & 409 &  8.66 & $\le -$4.00 &  34.2 &  11.3 \\
 546 & 416 &  1.00 & $\le -$4.00 &   5.6 &   1.2 \\
 547 & 408 &  1.22 & $-$0.98 &   1.0 &   2.5 \\
 550 & 9047 &  1.25 & $\le -$4.00 & 138.3 &   2.1 \\
 551 & 411 &  1.67 & $\le -$4.00 &   9.4 &  \nodata \\
 553 & 412 &  1.01 & $\le -$4.00 &  81.7 &  $-$3.5 \\
 557 & 415 &  0.05 & $\le -$4.00 &  27.8 &   1.8 \\
 558 & 418 &  1.59 & $\le -$4.00 &   5.3 &  \nodata \\
 560 & \nodata &  0.11 & $\le -$4.00 &   3.4 &  \nodata \\
 561 & 413 &  3.63 & $\le -$4.00 &  12.0 &   1.0 \\
 562 & 9048 &  0.72 & $-$1.61 &   1.0 &  $-$5.6 \\
 565 & 417 &  3.52 & $\le -$4.00 &   3.0 &   0.0 \\
 566 & 422 &  5.06 & $\le -$4.00 &  57.7 &  \nodata \\
 567 & 421 &  2.02 & $\le -$4.00 &  10.8 &  $-$3.5 \\
 569 & 419 &  1.40 & $-$2.12 &   1.0 &  \nodata \\
 573 & 420 &  1.22 & $-$1.01 &   1.0 &  \nodata \\
 579 & 423 &  2.39 & $\le -$4.00 &  79.9 & $-$17.4 \\
 584 & 427 &  1.46 & $-$1.19 &   1.0 &  \nodata \\
 585 & 428 &  0.42 & $\le -$4.00 &   2.5 &   1.4 \\
 586 & 425 &  0.28 & $\le -$4.00 &  99.8 &  \nodata \\
 593 & \nodata &  0.95 & $\le -$4.00 &   4.2 &  \nodata \\
 597 & 429 &  6.02 & $\le -$4.00 &  20.1 &   4.5 \\
 600 & 432 &  0.98 & $\le -$4.00 &  45.4 &  \nodata \\
 602 & 431 &  0.38 & $\le -$4.00 &  10.9 &  \nodata \\
 604 & 3093 &  0.00 & $-$1.88 &   1.7 &  \nodata \\
 605 & 3099 & $-$0.02 & $-$1.53 &   1.0 &   0.0 \\
 606 & 434 &  0.12 & $\le -$4.00 &   2.3 &  \nodata \\
 612 & 435a &  1.84 & $\le -$4.00 &   1.5 &  \nodata \\
 616 & 437 &  2.72 & $\le -$4.00 &   3.0 &   1.4 \\
 618 & 438 &  1.45 & $-$3.10 &   8.5 &  \nodata \\
 621 & \nodata &  2.84 & $\le -$4.00 &   7.1 &  \nodata \\
 622 & 9063 &  0.86 & $\le -$4.00 &  30.0 &  \nodata \\
 625 & \nodata &  2.01 & $\le -$4.00 &   8.8 &  \nodata \\
 626 & 439 &  2.53 & $\le -$4.00 &  19.3 &   0.7 \\
 631 & 441 &  2.14 & $\le -$4.00 &   4.2 &   0.7 \\
 632 & \nodata &  1.34 & $-$0.50 &   1.0 &  \nodata \\
 634 & 443 &  1.79 & $\le -$4.00 &  36.2 &  $-$1.5 \\
 636 & 447 &  2.05 & $-$1.67 &   1.0 &  \nodata \\
 638 & 442 &  1.28 & $-$0.22 &   1.0 &  $-$1.4 \\
 643 & 446 &  0.02 & $-$0.40 &   1.0 &   0.0 \\
 645 & 445a &  0.87 & $\le -$4.00 &   8.2 &   4.4 \\
 648 & 448 &  0.53 & $\le -$4.00 &   3.5 &  \nodata \\
 649 & 9069 & $-$0.09 & $\le -$4.00 &  61.9 &   0.0 \\
 651 & 444 &  1.00 & $-$0.30 &   1.0 &  \nodata \\
 653 & 450 &  1.87 & $\le -$4.00 &   6.4 &   0.8 \\
 658 & 451 &  0.32 & $\le -$4.00 &   6.5 &  $-$8.4 \\
 663 & 452 &  0.49 & $\le -$4.00 &   3.2 &  \nodata \\
 665 & \nodata &  0.90 & $\le -$4.00 &   1.6 &  \nodata \\
 666 & 453 &  0.60 & $\le -$4.00 &   4.5 &  \nodata \\
 669 & 457 &  4.69 & $\le -$4.00 &   9.7 &  \nodata \\
 670 & 454 &  1.00 & $\le -$4.00 &  24.9 &  $-$1.0 \\
 672 & 456 &  0.96 & $\le -$4.00 &   8.1 &  \nodata \\
 682 & 463 & 12.94 & $\le -$4.00 &  80.8 &  \nodata \\
 684 & 464 &  2.64 & $-$2.39 &   1.0 &  \nodata \\
 685 & 461 &  0.51 & $-$3.70 &  14.2 &  \nodata \\
 686 & 9092 &  0.29 & $-$0.58 &   1.0 &  $-$3.5 \\
 688 & 467 &  0.89 & $-$2.37 &   4.4 &   0.7 \\
 689 & 468 &  1.53 & $\le -$4.00 &   8.0 &   1.0 \\
 694 & 462 &  0.28 & $-$2.62 &  10.3 &   0.0 \\
 695 & 469 &  4.57 & $\le -$4.00 &   2.2 &   0.0 \\
 697 & 470 &  3.54 & $\le -$4.00 &   5.2 &   6.1 \\
 700 & 5177 &  2.82 & $\le -$4.00 &   3.8 &  \nodata \\
 701 & 471 &  9.36 & $\le -$4.00 &   3.9 &  \nodata \\
 703 & \nodata &  0.60 & $-$0.02 &   1.0 &  \nodata \\
 705 & 473 &  2.11 & $\le -$4.00 &  93.7 & $-$31.3 \\
 707 & 476 &  0.97 & $\le -$4.00 &   9.7 &   1.6 \\
 708 & 475 &  0.52 & $\le -$4.00 &  32.8 &  \nodata \\
 710 & 3103 &  0.51 & $\le -$4.00 &  47.8 &  \nodata \\
 711 & 477 &  1.19 & $\le -$4.00 &  12.2 &   1.3 \\
 717 & 5178 &  0.81 & $-$0.44 &   1.0 &  \nodata \\
 718 & 478 &  4.82 & $\le -$4.00 &  12.8 &  \nodata \\
 720 & 3032 & $-$0.20 & $\le -$4.00 &  21.9 &  \nodata \\
 724 & 479 &  1.44 & $\le -$4.00 &   2.5 &   2.0 \\
 726 & 480 &  0.70 & $\le -$4.00 &   2.4 &  \nodata \\
 727 & 9108 &  0.51 & $-$0.03 &   1.0 &  $-$9.3 \\
 728 & 482 &  1.63 & $\le -$4.00 &  36.2 & $-$50.3 \\
 731 & 481 & $-$0.42 & $-$0.45 &   1.0 &   0.0 \\
 737 & 485 &  7.07 & $\le -$4.00 &  18.3 &   0.8 \\
 738 & 486 &  0.00 & $-$0.03 &   1.0 &   0.0 \\
 739 & 490 &  0.52 & $\le -$4.00 &  31.5 &  $-$1.2 \\
 742 & 492 &  0.68 & $\le -$4.00 &  26.4 &  $-$0.9 \\
 743 & 494 &  0.28 & $-$1.10 &   2.3 &   0.6 \\
 746 & 488a &  1.34 & $-$3.40 &   2.6 &  $-$6.5 \\
 748 & 9118 &  0.68 & $-$0.35 &   1.0 & $-$53.0 \\
 750 & 502 &  4.64 & $\le -$4.00 &   4.7 &   0.0 \\
 752 & 501 &  1.24 & $\le -$4.00 & 132.6 &   1.1 \\
 753 & 487 & 10.69 & $\le -$4.00 &  33.8 &   1.8 \\
 754 & 493 &  0.31 & $\le -$4.00 &   6.8 &  \nodata \\
 756 & 496 &  4.47 & $-$3.22 &   1.9 & $-$10.8 \\
 758 & 499 &  3.00 & $\le -$4.00 &  11.2 & $-$12.3 \\
 761 & 504b &  0.13 & $\le -$4.00 &  18.3 &  \nodata \\
 762 & 497 &  1.26 & $\le -$4.00 &  67.6 &  \nodata \\
 763 & 500 &  0.61 & $-$0.84 &   1.0 &   0.5 \\
 768 & 503 &  0.11 & $\le -$4.00 &   4.4 &  $-$6.1 \\
 770 & 505 &  1.03 & $\le -$4.00 &  11.2 & $-$13.7 \\
 771 & 9124 &  1.21 & $\le -$4.00 &   3.4 &   0.0 \\
 773 & 3016 &  0.79 & $\le -$4.00 &  23.2 &  \nodata \\
 776 & 9132 & $-$0.15 & $\le -$4.00 &   2.6 &  $-$7.8 \\
 778 & 1863a &  9.70 & $-$1.16 &   1.0 &  \nodata \\
 782 & 9135 & $-$0.26 & $\le -$4.00 &   2.9 &  \nodata \\
 783 & 507 &  0.50 & $\le -$4.00 &  14.3 &  $-$4.0 \\
 789 & 509a &  0.92 & $\le -$4.00 & 121.5 &  \nodata \\
 790 & 3117 &  1.33 & $\le -$4.00 &  18.9 &  \nodata \\
 791 & 517 &  0.62 & $-$1.27 &   1.0 &   0.0 \\
 792 & 518 &  0.79 & $-$0.06 &   1.0 &   0.0 \\
 794 & 508 & 27.47 & $\le -$4.00 &  11.1 &   2.3 \\
 795 & 5056 & $-$0.07 & $\le -$4.00 &  11.4 &  \nodata \\
 796 & 510 &  1.15 & $-$3.00 &   3.3 &  $-$0.4 \\
 800 & 513 &  0.28 & $-$0.85 &   3.8 &  $-$1.7 \\
 801 & 515 &  2.85 & $\le -$4.00 &  14.0 &  $-$1.2 \\
 803 & 516 &  1.69 & $\le -$4.00 &  90.8 &   1.8 \\
 807 & 519a &  0.16 & $\le -$4.00 &  23.7 &   1.0 \\
 813 & 521 &  0.82 & $\le -$4.00 &  28.4 &   0.0 \\
 814 & 9147 &  1.86 & $\le -$4.00 &   9.3 &  $-$5.1 \\
 816 & 522 &  1.14 & $\le -$4.00 &  42.5 &  \nodata \\
 817 & 520 &  7.62 & $\le -$4.00 &  30.2 & $-$10.6 \\
 821 & 527 &  1.55 & $\le -$4.00 &  12.8 &   0.8 \\
 822 & 3031 &  0.61 & $-$0.11 &   1.0 &  \nodata \\
 825 & 9151 &  1.25 & $\le -$4.00 &  45.4 &  $-$6.2 \\
 826 & 524 &  1.47 & $\le -$4.00 &  19.9 &  \nodata \\
 830 & 523 &  0.20 & $-$3.40 &   1.8 &   1.7 \\
 832 & 529 & $-$1.50 & $\le -$4.00 &   2.5 &  \nodata \\
 839 & 3055 &  0.72 & $-$2.14 &   4.3 &  \nodata \\
 844 & 532 &  1.58 & $-$0.96 &   1.0 &  \nodata \\
 845 & 533 &  1.37 & $\le -$4.00 &   2.3 &   1.8 \\
 847 & 534 &  4.31 & $-$2.00 &   1.0 & $-$69.5 \\
 848 & 535 &  0.95 & $\le -$4.00 & 117.3 &   0.0 \\
 849 & \nodata &  0.91 & $-$3.52 &   8.8 &  \nodata \\
 851 & 528 &  0.52 & $-$1.34 &   3.8 &  \nodata \\
 853 & 9163 &  1.68 & $-$0.38 &   1.0 &  $-$4.6 \\
 854 & 536 &  0.45 & $\le -$4.00 &  22.7 &   1.7 \\
 855 & 538 &  4.56 & $\le -$4.00 &   3.0 &   1.1 \\
 856 & 537 &  1.45 & $\le -$4.00 &   2.8 &  $-$9.3 \\
 857 & 543 &  2.27 & $\le -$4.00 &  20.1 &  \nodata \\
 862 & 539 &  0.15 & $\le -$4.00 &   2.9 &  \nodata \\
 863 & 540 &  1.04 & $-$0.26 &   1.0 &  \nodata \\
 864 & 546 &  0.57 & $-$1.09 &   3.4 &  \nodata \\
 865 & 542 &  5.55 & $\le -$4.00 &  16.7 &  $-$1.0 \\
 867 & 544 &  2.02 & $\le -$4.00 &  15.0 &   1.6 \\
 870 & 9171 &  2.75 & $\le -$4.00 &   3.8 & $-$11.1 \\
 871 & 547 &  1.89 & $\le -$4.00 &   9.4 &   0.0 \\
 876 & 548 &  3.35 & $-$2.60 &   6.4 &  $-$3.6 \\
 881 & 551a &  2.22 & $\le -$4.00 &  80.3 &   2.0 \\
 882 & 555 &  0.50 & $-$0.70 &   1.0 &  $-$1.4 \\
 883 & 5047 & $-$0.03 & $-$0.40 &   1.0 &  \nodata \\
 887 & 9176 &  0.45 & $-$1.60 &   1.0 &  \nodata \\
 888 & 9179 &  1.69 & $\le -$4.00 &   6.7 &  $-$1.9 \\
 889 & 556 &  0.54 & $\le -$4.00 &   8.9 &  \nodata \\
 891 & 550 &  2.33 & $\le -$4.00 &  49.2 &   1.8 \\
 899 & 553a &  1.67 & $\le -$4.00 &  11.4 &  $-$0.7 \\
 901 & 557 & 28.29 & $\le -$4.00 &   2.4 &  $-$0.8 \\
 904 & \nodata &  1.81 & $\le -$4.00 &  11.0 &  \nodata \\
 907 & 565 & 10.88 & $\le -$4.00 &   5.9 &   0.0 \\
 914 & 559 &  1.41 & $-$3.22 &   6.9 &  $-$5.1 \\
 917 & 561 &  5.11 & $\le -$4.00 & 171.1 &  \nodata \\
 921 & 9201 &  4.68 & $-$3.40 &  11.4 &  $-$3.0 \\
 922 & 562 &  0.87 & $\le -$4.00 &   6.4 &  $-$2.2 \\
 924 & 9199 &  0.77 & $\le -$4.00 &  12.4 &  \nodata \\
 926 & 571 & 11.01 & $\le -$4.00 &   1.0 &  $-$3.3 \\
 927 & 560 &  3.26 & $\le -$4.00 &  29.0 &  \nodata \\
 928 & 563 &  0.33 & $\le -$4.00 &  25.9 &  \nodata \\
 929 & 564 & $-$0.47 & $\le -$4.00 &  58.7 &  \nodata \\
 932 & 567 & $-$0.01 & $\le -$4.00 &  12.1 &  $-$1.7 \\
 935 & 9204 &  2.06 & $-$0.97 &   1.0 & $-$14.1 \\
 936 & 569 &  1.54 & $\le -$4.00 &   5.2 &  \nodata \\
 937 & 570a &  4.55 & $\le -$4.00 &   2.8 &  \nodata \\
 938 & 573 &  1.79 & $\le -$4.00 & 136.4 &   0.0 \\
 939 & 566 &  0.53 & $\le -$4.00 &  11.0 &  \nodata \\
 942 & 9209 &  0.39 & $\le -$4.00 &  13.0 &   0.0 \\
 943 & 575 &  1.16 & $\le -$4.00 &  13.7 &  \nodata \\
 945 & 576 &  1.43 & $\le -$4.00 &  11.9 &   0.0 \\
 946 & 577 &  0.39 & $\le -$4.00 &  16.1 &   1.0 \\
 947 & 568 & $-$0.10 & $-$0.72 &   1.0 &   1.0 \\
 948 & 580 &  0.83 & $\le -$4.00 &   3.4 &  $-$7.7 \\
 949 & 572 & $-$0.05 & $\le -$4.00 &  15.3 &  \nodata \\
 954 & 9220 &  0.77 & $-$0.75 &   1.0 &  $-$1.8 \\
 955 & 582 &  1.80 & $-$0.36 &   6.8 &  $-$4.2 \\
 956 & 581 &  2.87 & $\le -$4.00 &   1.8 &  \nodata \\
 957 & 9213 &  0.63 & $-$0.90 &   1.0 &  \nodata \\
 958 & 583 & $-$1.17 & $-$1.01 &   1.0 &   3.1 \\
 960 & 590 &  0.21 & $\le -$4.00 & 441.0 &   0.0 \\
 962 & 584 &  0.70 & $-$1.34 &   3.6 &   0.0 \\
 963 & 586 &  1.66 & $\le -$4.00 &  83.1 &  \nodata \\
 965 & 589 & $-$0.13 & $\le -$4.00 &   1.4 &  $-$0.6 \\
 969 & 591 &  1.22 & $\le -$4.00 &   6.9 &  \nodata \\
 970 & 3140 &  0.20 & $\le -$4.00 &  50.1 &  \nodata \\
 971 & 585 &  1.13 & $\le -$4.00 &  21.9 &   1.8 \\
 972 & 9228 &  0.21 & $\le -$4.00 &  31.3 &  $-$2.0 \\
 974 & 592 &  5.06 & $-$1.29 &   2.1 &   0.0 \\
 976 & 594 &  1.54 & $\le -$4.00 & 193.4 &   0.0 \\
 982 & 593 &  1.15 & $\le -$4.00 &   6.3 &   1.8 \\
 983 & 10541 &  1.86 & $-$2.26 &   1.0 &  \nodata \\
 985 & 595 &  1.49 & $\le -$4.00 &  11.6 &  \nodata \\
 986 & 596 & 14.88 & $\le -$4.00 &  11.9 &  $-$7.7 \\
 987 & 9232 &  1.35 & $-$0.13 &   1.0 & $-$17.1 \\
 989 & 10543 &  0.98 & $-$0.16 &   1.0 &  \nodata \\
 990 & 5058 &  2.97 & $\le -$4.00 &   2.3 &  \nodata \\
 992 & 600 &  0.24 & $\le -$4.00 &   3.7 &  \nodata \\
 993 & 598a & $-$0.68 & $\le -$4.00 &   7.2 &  $-$0.3 \\
 994 & 597 & 14.28 & $\le -$4.00 &  37.7 &  \nodata \\
 995 & 599 &  7.07 & $-$1.29 &   1.0 &   2.8 \\
 997 & 603 &  4.78 & $\le -$4.00 &  39.5 &  \nodata \\
1000 & 604 &  1.94 & $\le -$4.00 &  15.6 & $-$19.0 \\
1001 & 9238 &  1.36 & $-$3.52 &   2.8 & $-$17.9 \\
1002 & 601 & $-$0.21 & $\le -$4.00 &  28.4 &   2.5 \\
1006 & 605 &  5.29 & $\le -$4.00 &  13.1 &  $-$2.3 \\
1007 & 606 &  2.49 & $\le -$4.00 &  19.8 &  \nodata \\
1008 & 9243 &  3.60 & $\le -$4.00 &   1.3 &  $-$4.2 \\
1011 & 607 &  0.59 & $\le -$4.00 &   1.8 &   2.3 \\
1012 & 9245 &  1.16 & $\le -$4.00 &   9.7 & $-$15.3 \\
1019 & 610 &  1.22 & $\le -$4.00 &  14.1 &  $-$5.8 \\
1023 & 9250 &  6.02 & $\le -$4.00 &   3.8 &   0.6 \\
1024 & 615 &  0.04 & $-$0.13 &   1.0 &  $-$4.5 \\
1025 & 5147 &  1.49 & $\le -$4.00 &   2.2 &  \nodata \\
1026 & 612 &  6.39 & $-$0.45 &   1.0 & $-$24.7 \\
1027 & 10553 & $-$0.04 & $\le -$4.00 &  20.5 &  \nodata \\
1028 & 613 &  0.92 & $\le -$4.00 &  12.2 &  \nodata \\
1032 & 617 &  0.79 & $-$0.36 &   1.0 &  21.4 \\
1034 & 5041 &  1.36 & $-$1.66 &   6.1 &  \nodata \\
1036 & 616 &  0.84 & $-$0.98 &   1.0 &  $-$3.3 \\
1038 & 621 &  2.26 & $-$0.28 &   1.0 &  \nodata \\
1044 & 620 &  3.28 & $\le -$4.00 &   8.6 &  $-$2.8 \\
1045 & 622 &  0.83 & $\le -$4.00 &  23.9 &  $-$7.0 \\
1047 & 619 &  0.11 & $-$0.13 &   1.0 &   0.0 \\
1056 & 624 &  2.26 & $-$2.89 &   3.3 &   0.0 \\
1058 & 625 &  0.05 & $\le -$4.00 &   4.0 &   6.1 \\
1059 & 632 &  3.28 & $-$3.22 &   2.6 &   0.0 \\
1060 & 628 &  1.12 & $\le -$4.00 &   2.2 &  \nodata \\
1061 & 5066 &  0.68 & $\le -$4.00 &  62.4 &  \nodata \\
1064 & 9256 &  3.99 & $-$0.07 &   1.0 & $-$10.4 \\
1065 & 633 &  2.59 & $-$2.72 &   1.0 &  \nodata \\
1070 & 634 &  0.77 & $\le -$4.00 &  56.8 &  \nodata \\
1071 & 631 &  4.23 & $\le -$4.00 &   6.2 &   1.6 \\
1076 & 636 & 18.35 & $\le -$4.00 &  17.0 &  \nodata \\
1077 & 638 &  2.61 & $-$2.70 &   1.0 &  \nodata \\
1080 & 629 &  3.06 & $\le -$4.00 &  25.0 & $-$16.9 \\
1083 & 637 &  2.19 & $\le -$4.00 &  11.7 &  \nodata \\
1085 & 3075 &  0.95 & $\le -$4.00 &  10.3 &  \nodata \\
1087 & 641 &  1.79 & $\le -$4.00 &   8.4 &   0.0 \\
1088 & 642 &  0.79 & $-$0.14 &   1.0 &   0.0 \\
1089 & 635 &  0.10 & $\le -$4.00 &   1.6 &   0.0 \\
1090 & 640 &  6.37 & $\le -$4.00 &   6.2 &  \nodata \\
1093 & 649 &  1.16 & $-$0.26 &   1.0 &   0.0 \\
1095 & 5119 &  0.83 & $\le -$4.00 &  37.6 &  $-$5.0 \\
1096 & 647 & 26.52 & $\le -$4.00 & 220.9 &  $-$5.0 \\
1097 & 643 &  4.41 & $\le -$4.00 &   6.0 &   1.0 \\
1100 & 645 &  1.56 & $\le -$4.00 &   3.9 &  13.6 \\
1101 & 648a & $-$0.36 & $\le -$4.00 &  24.1 &   0.9 \\
1102 & 653 &  0.05 & $-$2.03 &   1.0 &   0.0 \\
1103 & 651 &  0.77 & $\le -$4.00 &  16.7 &  \nodata \\
1104 & 650 &  0.95 & $\le -$4.00 &  10.5 &  \nodata \\
1107 & 652 &  1.09 & $-$0.87 &   1.0 &  $-$8.2 \\
1110 & 657 &  1.06 & $\le -$4.00 &  37.0 &   2.0 \\
1111 & 659 &  2.79 & $\le -$4.00 &  33.4 &  \nodata \\
1112 & 658 &  0.84 & $\le -$4.00 &   6.3 &  $-$7.6 \\
1113 & 10582 &  1.22 & $-$0.75 &   1.0 &  \nodata \\
1114 & 655 &  1.47 & $\le -$4.00 &  20.4 &  $-$1.5 \\
1116 & 660 & $-$0.14 & $\le -$4.00 &   5.6 &  \nodata \\
1117 & 661 &  0.97 & $\le -$4.00 &   9.0 &   2.1 \\
1118 & 3106 &  2.02 & $-$0.63 &   1.0 &  \nodata \\
1120 & 662 &  3.89 & $-$3.52 &   8.2 &  $-$1.0 \\
1121 & 663 & $-$0.24 & $-$3.30 &   2.7 &   2.4 \\
1122 & 665 &  1.23 & $\le -$4.00 & 500.7 & $-$10.1 \\
1124 & 9271 &  0.20 & $\le -$4.00 &   3.9 &  $-$4.4 \\
1127 & 664 &  6.84 & $\le -$4.00 &  49.3 &   1.4 \\
1130 & 669 &  1.94 & $\le -$4.00 &   2.0 &   1.8 \\
1131 & 9272 &  1.46 & $\le -$4.00 &  16.3 &  \nodata \\
1132 & 672 &  2.75 & $\le -$4.00 & 105.5 &   3.0 \\
1134 & 673 &  0.92 & $\le -$4.00 &  26.6 &   1.3 \\
1135 & 668 &  1.15 & $\le -$4.00 &   3.6 &  \nodata \\
1138 & 9277 &  0.27 & $-$1.57 &   1.0 &   0.0 \\
1139 & 674 & 10.52 & $\le -$4.00 &   9.2 &   0.0 \\
1140 & 9278 &  1.01 & $\le -$4.00 &  72.2 &  \nodata \\
1141 & 9280 &  2.15 & $\le -$4.00 &   5.1 &  $-$4.6 \\
1143 & 678 & 13.85 & $\le -$4.00 &   7.1 &  \nodata \\
1144 & 680 & $-$0.40 & $-$0.46 &   1.0 &  \nodata \\
1148 & 9283 &  1.48 & $-$1.29 &   1.0 &  $-$6.8 \\
1150 & 679 &  1.10 & $-$0.54 &   1.0 &  $-$2.0 \\
1151 & 683 &  3.35 & $\le -$4.00 &   6.6 &   1.9 \\
1154 & 684 & $-$0.43 & $\le -$4.00 &  12.7 &   0.0 \\
1155 & 685 &  1.08 & $-$0.36 &   1.0 &  $-$1.3 \\
1158 & 687a &  3.84 & $\le -$4.00 &  20.1 &   0.0 \\
1161 & 690 &  2.30 & $\le -$4.00 &   9.3 &   1.5 \\
1162 & 688 &  0.68 & $-$3.22 &   8.7 &  \nodata \\
1163 & 689 &  1.56 & $-$0.99 &   1.0 &   0.0 \\
1166 & 693 &  1.00 & $-$2.29 &   6.5 &  $-$2.9 \\
1167 & 694 &  1.93 & $\le -$4.00 &   3.8 &  $-$3.8 \\
1169 & 695 & $-$0.09 & $\le -$4.00 &  36.6 &   0.6 \\
1171 & 697 &  1.32 & $\le -$4.00 &  16.0 &  \nodata \\
1172 & 9293 & $-$0.14 & $\le -$4.00 &  44.2 &  \nodata \\
1174 & 698 &  2.91 & $-$2.43 &   1.5 &  \nodata \\
1177 & 692 & $-$0.64 & $\le -$4.00 &   1.7 &   1.6 \\
1178 & 691 &  0.35 & $\le -$4.00 &  13.5 &   0.0 \\
1182 & 702 & $-$0.22 & $-$0.43 &   1.0 &   3.1 \\
1184 & 5179 &  1.81 & $\le -$4.00 &  13.6 &  \nodata \\
1185 & 3116 &  0.67 & $-$0.47 &   1.0 &  \nodata \\
1186 & 704 &  1.33 & $\le -$4.00 &   8.6 &  \nodata \\
1187 & 3121 &  0.28 & $-$0.33 &   1.0 &  \nodata \\
1190 & 9298 &  1.21 & $-$3.00 &   3.8 &  \nodata \\
1191 & 9297 & $-$0.07 & $\le -$4.00 & 160.9 &  $-$5.3 \\
1193 & 707 & 10.08 & $\le -$4.00 &  12.3 &  \nodata \\
1195 & 3034 &  0.16 & $-$0.69 &   1.0 &  \nodata \\
1199 & 706 & $-$0.14 & $\le -$4.00 &   1.8 &  \nodata \\
1201 & 9299 &  2.23 &  0.00 &   1.0 &  \nodata \\
1202 & 709 &  0.33 & $\le -$4.00 &   7.9 &   2.8 \\
1204 & 711 &  1.01 & $-$2.06 &   1.0 &   0.0 \\
1205 & 710 &  4.73 & $-$0.68 &   1.0 &  $-$2.1 \\
1206 & 716 &  3.14 & $\le -$4.00 &   8.1 &  $-$2.7 \\
1207 & 715 &  1.21 & $\le -$4.00 &  28.9 &   0.0 \\
1210 & 712 &  6.51 & $\le -$4.00 &   6.3 &   0.8 \\
1212 & 713 &  1.78 & $\le -$4.00 &  12.8 &  \nodata \\
1213 & 718 &  3.28 & $\le -$4.00 &   3.8 &  \nodata \\
1214 & 720 & $-$0.19 & $\le -$4.00 &   7.5 &  \nodata \\
1216 & 721 &  5.34 & $\le -$4.00 &  13.4 &  $-$3.8 \\
1217 & 5007 & $-$0.31 & $-$0.16 &   1.0 &  \nodata \\
1218 & 714 &  0.48 & $-$0.26 &   1.0 &  \nodata \\
1219 & 717 &  2.61 & $\le -$4.00 &   9.1 &   0.0 \\
1222 & 3023 &  2.11 & $-$0.76 &   1.0 &  \nodata \\
1231 & 723 &  7.81 & $\le -$4.00 &  18.4 &  $-$1.1 \\
1233 & 727 &  4.96 & $\le -$4.00 &  10.4 &   0.8 \\
1236 & 728 &  6.58 & $\le -$4.00 &  40.2 &   0.9 \\
1239 & 5009 &  0.07 & $\le -$4.00 &   3.7 &  \nodata \\
1241 & 3105 &  3.37 & $-$0.13 &   1.0 &  \nodata \\
1242 & 730 &  0.06 & $-$1.03 &   1.0 &  \nodata \\
1243 & 734 &  0.18 & $-$0.27 &   1.0 &  \nodata \\
1245 & 732 &  1.32 & $\le -$4.00 &  57.8 & $-$15.8 \\
1246 & 735 &  3.71 & $\le -$4.00 & 207.2 &   0.0 \\
1247 & 5027 &  1.11 & $-$0.18 &   1.0 &  \nodata \\
1248 & 733 &  0.01 & $\le -$4.00 &  11.0 &   1.7 \\
1252 & 736 &  2.50 & $\le -$4.00 &  81.2 &   1.6 \\
1259 & 738 &  0.08 & $\le -$4.00 & 150.9 &  $-$0.3 \\
1260 & 740 &  1.57 & $-$0.93 &   1.0 &  $-$8.2 \\
1261 & 741 & $-$0.13 & $\le -$4.00 &  76.8 &  \nodata \\
1262 & 737 &  1.98 & $-$2.31 &   1.0 &  \nodata \\
1263 & 743 & 12.62 & $-$0.27 &   1.0 &  $-$2.0 \\
1264 & 739 &  6.89 & $\le -$4.00 &   8.5 &  $-$1.8 \\
1265 & 742 &  0.31 & $-$2.54 &   1.0 &  \nodata \\
1266 & 5131 &  0.88 &  0.00 &   1.0 &  \nodata \\
1267 & 9312 &  0.72 & $\le -$4.00 &  61.3 &  \nodata \\
1268 & 744 & $-$0.14 & $\le -$4.00 &  82.7 &  $-$0.9 \\
1269 & 747 & $-$0.33 & $\le -$4.00 &  11.4 &  \nodata \\
1271 & 746 &  0.24 & $\le -$4.00 &  29.7 &   0.6 \\
1275 & 3087 &  2.35 & $\le -$4.00 &   4.6 &  $-$4.7 \\
1276 & \nodata &  1.78 & $\le -$4.00 & 136.2 &  \nodata \\
1279 & 748a &  1.43 & $\le -$4.00 &  21.6 &  $-$1.4 \\
1280 & 9316 &  1.04 & $\le -$4.00 &  42.6 &   0.0 \\
1281 & 750 &  3.08 & $\le -$4.00 &   2.8 &   0.0 \\
1282 & 753 &  3.26 & $\le -$4.00 &   3.1 &  \nodata \\
1286 & 3101 & $-$0.34 & $-$1.29 &   1.0 &  \nodata \\
1287 & 10620 &  0.87 & $-$0.27 &   1.0 &  \nodata \\
1288 & 755 &  0.26 & $-$1.15 &   1.0 &   0.0 \\
1289 & 757 &  0.99 & $-$1.42 &   1.0 &  \nodata \\
1290 & 756 &  1.72 & $\le -$4.00 & 146.0 &  $-$2.0 \\
1291 & 758 &  6.08 & $\le -$4.00 &   4.0 &  $-$2.0 \\
1292 & 751 &  8.05 & $\le -$4.00 & 146.7 &  $-$3.0 \\
1293 & 759 &  1.81 & $-$0.63 &   1.0 &  \nodata \\
1295 & 760a & 11.28 & $-$1.76 &   1.0 &  $-$8.5 \\
1299 & 761 &  0.19 & $\le -$4.00 &  15.5 &  \nodata \\
1302 & 762 & 12.33 & $\le -$4.00 &  19.4 &  $-$4.5 \\
1305 & 763 &  2.82 & $\le -$4.00 &  85.1 &   0.8 \\
1306 & 764 &  0.40 & $\le -$4.00 &   9.2 &  \nodata \\
1307 & 10624 &  2.59 & $\le -$4.00 & 100.4 &  \nodata \\
1309 & 766a &  1.96 & $\le -$4.00 &   7.9 &  \nodata \\
1311 & 769 &  1.53 & $\le -$4.00 &   4.1 &  \nodata \\
1312 & 768 &  3.79 & $\le -$4.00 &   1.5 & $-$14.8 \\
1314 & 770 &  0.71 & $\le -$4.00 &   5.6 &  $-$3.7 \\
1316 & 767 & 32.01 & $\le -$4.00 &  11.4 &  $-$1.8 \\
1317 & 773 &  0.33 & $-$0.87 &   1.0 &  \nodata \\
1319 & 3128 &  0.28 & $-$1.94 &   1.0 &  \nodata \\
1323 & 775 & $-$0.89 & $-$2.89 &   3.4 &   0.0 \\
1324 & 3095 &  4.71 & $\le -$4.00 & 111.7 &  \nodata \\
1326 & 777 &  0.98 & $-$3.70 &  10.2 &   2.5 \\
1327 & 777 &  1.00 & $\le -$4.00 &   6.9 &   2.5 \\
1328 & 776a &  6.50 & $\le -$4.00 &  28.3 &  \nodata \\
1329 & 3079 &  0.53 & $\le -$4.00 &  46.2 &  \nodata \\
1332 & 5059 &  0.49 & $-$0.10 &   1.0 &  \nodata \\
1333 & 787 & 25.02 & $\le -$4.00 &   5.2 &   0.6 \\
1334 & 788 & 14.80 & $-$0.17 &   8.4 &   0.0 \\
1335 & 782 &  5.40 & $\le -$4.00 &  22.6 &  $-$2.5 \\
1336 & 786 &  5.55 & $\le -$4.00 &  13.9 &   0.0 \\
1341 & 791 &  0.49 & $\le -$4.00 &  24.7 &  \nodata \\
1342 & 792 &  3.97 & $\le -$4.00 &   6.1 &   1.4 \\
1343 & 793 &  3.17 & $\le -$4.00 &  68.3 &  \nodata \\
1344 & 3069 &  2.03 & $\le -$4.00 &  15.0 &  \nodata \\
1345 & 3090 &  0.66 & $-$1.17 &   1.0 &   0.0 \\
1347 & 796 & 24.80 & $-$0.59 &   1.0 &  $-$9.3 \\
1350 & 795 &  3.28 & $\le -$4.00 &  12.7 &   1.7 \\
1354 & 799 &  1.90 & $\le -$4.00 &  11.1 &  \nodata \\
1355 & 798 &  4.99 & $\le -$4.00 &   4.7 &   0.0 \\
1356 & 3021 &  1.28 & $\le -$4.00 &   4.7 &  \nodata \\
1358 & 800 &  0.15 & $-$0.67 &   1.0 &   0.0 \\
1361 & 802 &  4.10 & $-$0.18 &   1.0 &  \nodata \\
1362 & 801 & $-$0.29 & $\le -$4.00 &   3.7 &  \nodata \\
1363 & 797 & 13.11 & $-$1.72 &   6.5 &  \nodata \\
1366 & 3027 & 10.37 & $-$1.29 &   1.0 &  \nodata \\
1370 & 10649 &  0.48 & $\le -$4.00 &   3.0 &  \nodata \\
1371 & 10653 & $-$0.63 & $-$2.24 &   1.0 &  \nodata \\
1372 & 807 & $-$0.02 & $-$3.70 &   9.4 &   2.1 \\
1373 & 806 &  0.69 & $-$1.28 &   1.0 &   2.5 \\
1374 & 805 &  1.50 & $\le -$4.00 &  46.7 &  \nodata \\
1378 & 812 &  0.38 & $-$0.87 &   1.0 &  \nodata \\
1380 & 810 &  2.16 & $\le -$4.00 &  19.3 &  \nodata \\
1382 & 811a & 17.77 & $\le -$4.00 &   9.1 &   0.0 \\
1384 & 813 &  4.24 & $\le -$4.00 &  20.0 &   1.9 \\
1385 & 10659 &  0.67 & $-$0.98 &   1.0 &  \nodata \\
1387 & 815 &  4.65 & $\le -$4.00 &  36.1 &   0.0 \\
1388 & 817 & $-$0.11 & $\le -$4.00 &  28.5 &  \nodata \\
1391 & 816 &  3.18 & $\le -$4.00 &   9.7 &   1.4 \\
1394 & 3077 &  0.59 & $-$0.90 &   1.0 &  \nodata \\
1396 & 3119 &  1.29 & $\le -$4.00 &  42.8 &  \nodata \\
1397 & \nodata &  1.60 & $-$0.01 &   1.0 &  \nodata \\
1398 & 823 &  0.69 & $\le -$4.00 &  80.1 &   0.0 \\
1400 & 819 &  0.73 & $-$2.66 &   4.5 &   1.3 \\
1403 & 824 & 18.43 & $-$0.24 &   1.0 & $-$13.4 \\
1404 & 821 &  3.51 & $\le -$4.00 &   6.9 &   0.0 \\
1407 & 825 & $-$0.10 & $-$2.41 &   2.2 &  $-$1.1 \\
1408 & 5035 &  0.96 & $\le -$4.00 &  13.8 &  \nodata \\
1409 & 826 & 11.03 & $\le -$4.00 &  38.6 &  $-$6.3 \\
1410 & 827 &  1.07 & $\le -$4.00 & 211.6 &   0.0 \\
1411 & 828 &  2.01 & $\le -$4.00 &  20.8 &   0.0 \\
1412 & 830 &  4.54 & $\le -$4.00 &   3.3 &   1.8 \\
1413 & 829 &  2.51 & $\le -$4.00 &   2.0 &   0.0 \\
1415 & 831 & $-$0.37 & $\le -$4.00 &   1.8 &  \nodata \\
1419 & 834 & 15.59 & $\le -$4.00 & 110.0 &  \nodata \\
1421 & 836 & 27.27 & $\le -$4.00 &  16.6 &   0.9 \\
1422 & 10672 &  0.20 & $-$0.04 &   1.0 &  \nodata \\
1423 & 838 &  1.69 & $\le -$4.00 &  10.9 &   0.0 \\
1424 & 837 &  6.89 & $\le -$4.00 &   3.4 &   1.2 \\
1425 & 3042 &  1.47 & $\le -$4.00 &   2.2 &  \nodata \\
1426 & 3097 &  1.05 & $-$3.05 &   4.7 &  \nodata \\
1429 & 839 &  6.17 & $\le -$4.00 &   4.2 &   1.2 \\
1431 & 3129 &  2.44 & $-$3.52 &   6.1 &  \nodata \\
1432 & 843 &  0.02 & $\le -$4.00 &   4.8 &   0.0 \\
1433 & 3019 &  3.23 & $\le -$4.00 &  86.4 &  \nodata \\
1434 & 845 &  0.91 & $\le -$4.00 &   1.9 & $-$12.2 \\
1435 & 844 &  7.96 & $-$1.11 &   1.0 &   1.3 \\
1438 & 848 &  2.20 & $\le -$4.00 &   7.9 &   2.1 \\
1440 & 849 &  0.56 & $-$1.49 &   1.6 &   0.9 \\
1443 & 852 &  3.11 & $\le -$4.00 &  63.2 &  \nodata \\
1444 & 850 & 27.83 & $\le -$4.00 &  96.7 &  $-$4.1 \\
1445 & 855 &  7.18 & $-$1.93 &   3.2 &   0.0 \\
1447 & 857 & $-$0.07 & $-$2.96 &  22.0 &  $-$4.2 \\
1448 & 860 &  6.52 & $\le -$4.00 &   2.8 &   2.1 \\
1449 & 854 &  8.90 & $\le -$4.00 &  78.0 &  \nodata \\
1450 & 10680 &  0.49 & $-$3.15 &   1.3 &  \nodata \\
1451 & 859 &  1.21 & $\le -$4.00 &   9.8 &  $-$2.6 \\
1454 & 863 & 13.82 & $\le -$4.00 &   2.8 &   0.0 \\
1455 & 861 &  0.30 & $\le -$4.00 &  79.2 &  \nodata \\
1456 & 3061 &  1.06 & $\le -$4.00 &  51.9 &  \nodata \\
1457 & 3092 &  2.44 & $-$1.17 &   1.0 &  \nodata \\
1458 & 865 &  0.04 & $-$0.67 &   1.0 &   1.5 \\
1459 & 862 &  0.88 & $-$0.58 &   1.0 &  $-$4.5 \\
1462 & 866 &  0.55 & $\le -$4.00 & 257.8 &  \nodata \\
1463 & 867 &  8.16 & $\le -$4.00 &  10.6 &  \nodata \\
1464 & 5071 & $-$0.06 & $\le -$4.00 &  22.8 &  \nodata \\
1466 & 868 &  7.74 & $\le -$4.00 &  22.8 &   1.4 \\
1470 & 5126 &  0.20 & $-$2.77 &   1.0 &  \nodata \\
1473 & 2085 &  0.06 & $\le -$4.00 &  11.7 &  \nodata \\
1474 & 872 &  1.95 & $\le -$4.00 &  33.8 &   0.4 \\
1475 & 870 &  2.61 & $-$2.57 &   3.8 &  \nodata \\
1477 & 10689 &  0.62 & $-$1.12 &   1.0 &  \nodata \\
1478 & 871 &  2.18 & $\le -$4.00 &  15.3 &  \nodata \\
1479 & 878 &  2.38 & $\le -$4.00 &   6.9 &   0.0 \\
1481 & 880 &  2.08 & $\le -$4.00 &  15.2 &   0.0 \\
1483 & 5156 &  1.38 & $-$0.40 &   1.0 &  \nodata \\
1484 & 877 &  0.17 & $\le -$4.00 &  12.3 &  \nodata \\
1485 & 879 &  3.96 & $\le -$4.00 &  52.0 &  $-$1.4 \\
1487 & 883 &  1.45 & $\le -$4.00 &  67.5 &  \nodata \\
1489 & 887 & $-$0.94 & $\le -$4.00 &  15.9 &  \nodata \\
1490 & 3014 &  0.28 & $-$2.42 &   1.5 &  \nodata \\
1492 & 890 &  1.39 & $\le -$4.00 &  10.3 &   1.6 \\
1495 & 891 & $-$0.07 & $-$2.68 &   6.6 &   0.0 \\
1497 & 5060 &  0.41 & $-$1.60 &   1.0 &  \nodata \\
1499 & 895 & 45.45 & $\le -$4.00 & 283.7 &  \nodata \\
1500 & 892 &  3.85 & $\le -$4.00 & 106.2 &   1.0 \\
1501 & 894 &  0.71 & $\le -$4.00 &   9.4 &  \nodata \\
1503 & 5076 &  0.64 & $\le -$4.00 &  31.3 &   0.0 \\
1507 & 898 & $-$0.25 & $\le -$4.00 &  17.3 &   0.0 \\
1508 & 897 &  0.98 & $-$0.84 &   1.0 &  \nodata \\
1511 & 901 &  1.01 & $\le -$4.00 &  10.6 &  \nodata \\
1512 & 902 &  1.25 & $\le -$4.00 &   9.4 &  \nodata \\
1514 & 905 &  0.40 & $-$1.43 &   1.0 &  \nodata \\
1516 & 907 &  5.77 & $\le -$4.00 &   9.1 &   1.8 \\
1520 & 910 &  0.00 & $-$0.35 &   1.0 &   1.8 \\
1521 & 911 & 16.86 & $\le -$4.00 &   4.0 &  \nodata \\
1522 & 912 &  1.71 & $-$0.44 &   1.7 &   1.2 \\
1523 & 5085 &  0.69 & $-$0.35 &   1.0 &  \nodata \\
1524 & 916 &  0.04 & $-$2.57 &   1.0 &  \nodata \\
1529 & 5083 &  1.17 & $\le -$4.00 &   7.9 &  \nodata \\
1531 & 3049 &  1.01 & $\le -$4.00 &   1.7 &  \nodata \\
1532 & 5067 & $-$0.55 & $-$0.60 &   1.0 &  \nodata \\
1537 & 928 & $-$0.04 & $-$0.84 &   1.0 &  \nodata \\
1539 & 930 &  0.58 & $\le -$4.00 &   8.9 &   5.2 \\
1540 & 929 & $-$0.64 & $-$0.42 &   1.0 &   0.0 \\
1542 & 10723 &  0.16 & $-$2.74 &   1.0 &  \nodata \\
1543 & 933 & $-$0.13 & $\le -$4.00 &   6.5 &   1.0 \\
1544 & 931 &  0.45 & $\le -$4.00 &  11.9 &   2.0 \\
1546 & 936 &  1.49 & $-$2.96 &  22.1 &   0.8 \\
1547 & 939 &  0.42 & $-$1.99 &   1.0 &   2.1 \\
1549 & 942 & $-$0.03 & $\le -$4.00 &  17.5 &   0.0 \\
1550 & 941 & $-$0.51 & $\le -$4.00 &   7.6 &   1.6 \\
1553 & 944 &  2.76 & $\le -$4.00 &  31.8 &   1.2 \\
1558 & 950 &  0.13 & $-$0.67 &   1.0 &  \nodata \\
1560 & 10744 &  2.63 & $\le -$4.00 &  18.9 &  \nodata \\
1563 & 951 &  6.16 & $-$0.39 &   1.0 &  $-$3.2 \\
1564 & 954 &  9.83 & $-$3.70 &   1.0 &   0.0 \\
1568 & 959 &  0.01 & $\le -$4.00 &  90.0 &  \nodata \\
1569 & 958 &  0.55 & $-$3.70 &   2.8 &  $-$0.4 \\
1570 & 962 & 17.22 & $\le -$4.00 &  51.9 &   1.8 \\
1571 & 965 &  0.25 & $\le -$4.00 &   6.4 &   0.0 \\
1572 & 964 &  1.55 & $\le -$4.00 &   3.5 &   0.0 \\
1573 & 3107 &  0.87 & $-$1.74 &   1.0 &  \nodata \\
1576 & 969 & 19.22 & $-$0.05 &   1.0 &  $-$2.7 \\
1585 & 980 & 23.78 & $\le -$4.00 &   2.1 &   0.0 \\
1588 & 10776 &  0.86 & $-$1.70 &   6.2 &  \nodata \\
1590 & 982 & 37.01 & $\le -$4.00 &   3.8 &  $-$1.2 \\
1591 & 981 &  1.81 & $\le -$4.00 &  15.1 &   1.7 \\
1592 & 984 &  0.86 & $\le -$4.00 &  17.5 &   0.7 \\
1594 & 990 &  0.49 & $\le -$4.00 &  35.6 &   0.6 \\
1595 & 998 &  5.74 & $\le -$4.00 &   6.0 &  \nodata \\
1599 & 1013 &  1.18 & $-$3.70 &   4.2 & $-$26.5 \\
1603 & 5097 &  0.92 & $\le -$4.00 &  21.9 &   0.0 \\
1604 & 10817 & 16.31 & $-$0.47 &   1.0 &  \nodata \\
1608 & 1026a & 26.17 & $\le -$4.00 &  35.8 &  $-$1.3 \\
1610 & 1030 &  4.82 & $-$2.03 &   1.0 &   1.5 \\
1611 & 1032 &  1.26 & $-$0.05 &   1.0 &   0.0 \\
1612 & 1034 &  0.38 & $\le -$4.00 &   7.7 &   0.4 \\
1613 & 1037 &  1.35 & $-$0.25 &   1.0 &   2.5 \\
\enddata
\tablenotetext{a}{Optical identification from \citet{getman05}.}
\tablenotetext{b}{Value of $J$ statistic; see \S \ref{optvar}.}
\tablenotetext{c}{KS test probability of X-ray variability; see \S \ref{xvar}.
Values below $-4.00$ have been truncated \citep{getman05}.}
\tablenotetext{d}{Ratio of maximum/minimum X-ray flux from BB analysis; see \S \ref{xvar}.}
\tablenotetext{e}{Equivalent width in \AA\ of the \ion{Ca}{2} IR triplet lines. 
Negative values indicate emission. See \citet{hill98}.}
\tablecomments{This table is available in its entirety in the electronic
version of the Journal. A portion is shown here for guidance regarding
its form and content.}
\end{deluxetable}

\clearpage

\begin{deluxetable}{llrr}
\tablecolumns{4}
\tablewidth{0pt}
\tabletypesize{\scriptsize}
\tablecaption{Contingency table of optical and X-ray 
variability in the study sample\label{var-contingency-table}}
\tablehead{
\colhead{Optical} & \colhead{} & \multicolumn{2}{c}{X-ray variability} \\
\colhead{variability} & \colhead{} & \colhead{Yes} & \colhead{No} 
}
\startdata
\cutinhead{Entire study sample ($N=814$)}
Yes & & 278 &  80 \\
No  & & 278 & 178 \\
\cutinhead{Actively accreting stars ($N=151$)}
Yes & &  60 &  25 \\
No  & &  39 &  27 \\
\cutinhead{Non-accreting stars ($N=145$)}
Yes & &  64 &   8 \\
No  & &  59 &  14 \\
\enddata
\end{deluxetable}

\clearpage

\clearpage

\clearpage

\begin{thebibliography}{}

\bibitem[Audard et al.(2007)]{audard07} Audard, M., Briggs, K., 
Grosso, N., Guedel, M., Scelsi, L., Bouvier, J., \& Telleschi, A.\ 2007, 
\aap, in press

\bibitem[Feigelson et al.(2003)]{feig03} Feigelson, E.~D., 
Gaffney, J.~A., Garmire, G., Hillenbrand, L.~A., Townsley, L.\ 2003, 
\apj, 584, 911 

\bibitem[Feigelson et al.(2006)]{feig06} Feigelson, E., Townsley,
L., Gudel, M., \& Stassun, K.\ 2006, Protostars \& Planets V, in press

\bibitem[Flaccomio et al.(2005)]{flac-spots} Flaccomio, E., 
Micela, G., Sciortino, S., Feigelson, E.~D., Herbst, W., Favata, F., 
Harnden, F.~R., \& Vrtilek, S.~D.\ 2005, \apjs, 160, 450

\bibitem[Flaccomio, Micela, \& Sciortino(2003b)]{flac-time} 
Flaccomio, E., Micela, G., \& Sciortino, S.\ 2003b, \aap, 402, 277 

\bibitem[Flaccomio et al.(2003)]{flac-basic} Flaccomio, E., 
Damiani, F., Micela, G., Sciortino, S., Harnden, F.~R., Murray, S.~S., \& 
Wolk, S.~J.\ 2003, \apj, 582, 398

\bibitem[Getman et al.(2005)]{getman05}
Getman, K.~V., et al.\ 2005, \apjs, 160, 319 

\bibitem[Glassgold, Feigelson, \& Montmerle(2000)]{glassgold00} 
Glassgold, A.~E., Feigelson, E.~D., \& Montmerle, T.\ 2000, Protostars and 
Planets IV, 429

\bibitem[Guenther et al.(1999)]{guenther99} Guenther, E.~W., 
Lehmann, H., Emerson, J.~P., \& Staude, J.\ 1999, \aap, 341, 768 

\bibitem[Herbst et al.(2002)]{herbst02} Herbst, W., 
Bailer-Jones, C.~A.~L., Mundt, R., Meisenheimer, K., \& Wackermann, R.\ 
2002, \aap, 396, 513

\bibitem[Herbst et al.(1994)]{herbst94} 
Herbst, W., Herbst, D.~K., Grossman, E.~J., \& Weinstein, D.\ 1994, \aj, 
108, 1906

\bibitem[Hillenbrand(1997)]{hill97} Hillenbrand, L.~A.\ 1997, 
\aj, 113, 1733

\bibitem[Hillenbrand et al.(1998)]{hill98} Hillenbrand, L.~A., 
Strom, S.~E., Calvet, N., Merrill, K.~M., Gatley, I., Makidon, R.~B., 
Meyer, M.~R., \& Skrutskie, M.~F.\ 1998, \aj, 116, 1816 

\bibitem[Jardine et al.(2006)]{jardine06} Jardine, M., Cameron, 
A.~C., Donati, J.-F., Gregory, S.~G., \& Wood, K.\ 2006, \mnras, 367, 917 

\bibitem[Jeffries(1999)]{jeffries99} Jeffries, R.D.\ 1999, ASP 
Conf.~Ser.~158: Solar and Stellar Activity:  Similarities and Differences, 75 

\bibitem[Johns-Krull, Valenti, \& Saar(2004)]{johnskrull04}
Johns-Krull, C.~M., Valenti, J.~A., \& Saar, S.~H.\ 2004, \apj, 617, 1204

\bibitem[Johns-Krull et al.(1999)]{johnskrull99} Johns-Krull, C.~M.,
Valenti, J.~A., \& Koresko, C.\ 1999, \apj, 516, 900

\bibitem[Johns-Krull \& Hatzes(1997)]{johnskrull97} Johns-Krull,
C.~M., \& Hatzes, A.~P.\ 1997, \apj, 487, 896

\bibitem[Kastner et al.(2002)]{kastner02} Kastner, J.~H., 
Huenemoerder, D.~P., Schulz, N.~S., Canizares, C.~R., \& Weintraub, D.~A.\ 
2002, \apj, 567, 434

\bibitem[Pallavicini et al.(1981)]{pallavicini81} Pallavicini, R., 
Golub, L., Rosner, R., Vaiana, G.~S., Ayres, T., \& Linsky, J.~L.\ 1981, 
\apj, 248, 279

\bibitem[Pizzolato et al.(2003)]{pizzolato03} Pizzolato, N., 
Maggio, A., Micela, G., Sciortino, S., \& Ventura, P.\ 2003, \aap, 397, 147

\bibitem[Preibisch et al.(2005)]{preibisch05} 
Preibisch, T., et al.\ 2005, \apjs, 160, 401

\bibitem[Randich(2000)]{randich00} Randich, S.\ 2000, ASP 
Conf.~Ser.~198: Stellar Clusters and Associations: Convection, Rotation, 
and Dynamos, 401

\bibitem[Rebull et al.(2006)]{rebull06} Rebull, L.~M., Stauffer,
J.~R., Ramirez, S.~V., Flaccomio, E., Sciortino, S., Micela, G.,
Strom, S.~E., \& Wolff, S.~C.\ 2006, \aj, 131, 2934

\bibitem[Rebull(2001)]{rebull01} Rebull, L.~M.\ 2001, \aj, 121, 1676

\bibitem[Rice \& Strassmeier(1996)]{rice96} Rice, J.~B., \&
Strassmeier, K.~G.\ 1996, \aap, 316, 164

\bibitem[Scargle(1998)]{scargle98} Scargle, J.~D.\ 1998, \apj, 504, 405

\bibitem[Schrijver \& Zwaan(2000)]{schrijver00} Schrijver, C.~J.,
\& Zwaan, C.\ 2000, Solar and stellar magnetic activity.~New York: 
Cambridge University Press, 2000.~(Cambridge astrophysics series; 34)

\bibitem[Sicilia-Aguilar et al.(2005)]{sicilia05}
Sicilia-Aguilar, A., et al.\ 2005, \aj, 129, 363

\bibitem[Stassun et al.(1999)]{stass99}
Stassun, K.~G., Mathieu, R.~D., Mazeh, T., \& Vrba, F.~J.\ 1999,
\aj, 117, 2941

\bibitem[Stassun et al.(2004)]{stass04}
Stassun, K.~G., Ardila, D.~R., Barsony, M., Basri, G., \& Mathieu,
R.~D.\ 2004, \aj, 127, 3537

\bibitem[Stassun et al.(2006)]{stass06}
Stassun, K.~G., van den Berg, M., Feigelson, E.~D., \& Flaccomio,
E.\ 2006, \apj, in press (Paper I)

\bibitem[Stelzer \& Schmitt(2004)]{stelzer04} Stelzer, B., \& 
Schmitt, J.~H.~M.~M.\ 2004, \aap, 418, 687 

\bibitem[Stetson(1996)]{stetson96}
Stetson, P.~B.\ 1996, \pasp, 108, 851

\end{thebibliography}
\end{document}